\documentclass[aps,prb,twocolumn,groupadress]{revtex4}
\usepackage{ulem}
\usepackage{graphicx}
\usepackage{hyperref}
\usepackage{amsmath,amssymb}
\usepackage{bm}
\usepackage{verbatim}

\def\be{\begin{equation}}
\def\ee{\end{equation}}
\def\ba{\begin{array}{lll}}
\def\ea{\end{array}}
\def\ber{\begin{eqnarray}}
\def\eer{\end{eqnarray}}
\def\ep{\epsilon}
\def\om{\omega}
\def\kv{{\bm k}}
\def\hb{\hbar}
\def\De{\Delta}

\begin{document}

\title{Unified Boltzmann-transport theory for the drag resistivity close to a second-order phase transition}

\author{M. P. Mink}
\affiliation{Institute for Theoretical Physics, Utrecht
University, Leuvenlaan 4, 3584 CE Utrecht, The Netherlands}

\author{H. T. C. Stoof}
\affiliation{Institute for Theoretical Physics, Utrecht
University, Leuvenlaan 4, 3584 CE Utrecht, The Netherlands}

\author{R. A. Duine}
\email{R.A.Duine@uu.nl}
\affiliation{Institute for Theoretical Physics, Utrecht
University, Leuvenlaan 4, 3584 CE Utrecht, The Netherlands}

\author{Marco Polini}
\affiliation{NEST, Istituto Nanoscienze-CNR and Scuola Normale Superiore, I-56126 Pisa, Italy}

\author{G. Vignale}
\affiliation{Department of Physics and Astronomy, University of Missouri, Columbia, Missouri 65211, USA}

\date{\today}

\begin{abstract}
We present a unified Boltzmann-transport theory for the drag resistivity $\rho_{\rm D}$ in two-component systems close to a second-order phase transition. We find general expressions for $\rho_{\rm D}$ in two and three spatial dimensions, for arbitrary population and mass imbalance, for particle-
and hole-like bands, and show how to incorporate, at the Gaussian level, the effect of fluctuations close to a phase transition. We find that the proximity to the phase transition enhances the drag resistivity upon approaching the critical temperature from above, and we qualitatively derive the temperature dependence of this enhancement for various cases. In addition, we present numerical results for two concrete experimental systems: i) three-dimensional cold atomic Fermi gases close to a Stoner transition and ii) two-dimensional spatially-separated electron and hole systems in semiconductor double quantum wells.
\end{abstract}

\maketitle
\vskip2pc


%
\section{Introduction}
\label{sec:int}
The behavior of linear-response coefficients close to a phase transition has a long history as an interesting field of study. Two prime examples are the vanishing of the resistivity at the superconducting phase transition and the divergence of the magnetic susceptibility at a ferromagnetic transition. Indeed, the behavior of these coefficients is often the most important experimental signature in assessing whether or not the system has reached an ordered state. For example,  deviations from the standard low temperature $T^2$ dependence of the resistivity of a metal may indicate non-Fermi-liquid behavior that could result from a phase transition to a symmetry-broken state.

However, in electronic condensed-matter systems, the theoretical calculation of these coefficients is often very difficult, due to many competing phenomena: electron-phonon coupling, presence of impurities, localization, and Coulomb interactions between the carriers.  The prime advantage of a drag experiment is that it singles out the effect of Coulomb interactions on a transport coefficient: consider a system consisting of two layers, separated by a barrier so that tunneling between the layers is absent. In a drag experiment in this bilayer system, a current is driven through one of the layers, denoted as the drive layer. Due to momentum transfer, the carriers in the other (the passive) layer are dragged along and a voltage drop over the passive layer is observed. The drag resistance is defined as the ratio between the current in the active or drive layer and the voltage drop over the passive layer. Due to the spatial separation of the layers, the Coulomb interaction between carriers in both layers can be singled out as solely responsible for this effect, which for this reason is called ``Coulomb drag". In cold two-component Fermi gases a similar phenomenon can be observed, where now the two (hyperfine) spin species play the role of the carriers in the two layers. When a cloud of atoms with one spin state moves relative to another, interactions lead to momentum transfer and the second spin species is also set into motion. This phenomenon is called spin drag.

Coulomb drag was first observed in 1990 \cite{gramilla} for electron-electron bilayers and later also for electron-hole bilayers, \cite{sivan} for a review see Ref.~[\onlinecite{rojo}]. In electron-hole bilayers the electrons in one layer and holes in the other can form excitons, which are expected to condense for low enough temperatures. The behavior of the drag resistivity in this condensed state was first studied in Ref.~[\onlinecite{vignale96}], and  its enhancement above the critical temperature was calculated in Ref.~[\onlinecite{hu}]. An enhancement of the drag resistivity has been measured experimentally, \cite{ehexp} although exciton condensation has not been confirmed. Also for a topological insulator thin film, an enhancement of the drag resistivity upon approaching the critical temperature for (in this case topological) exciton condensation has recently been predicted. \cite{mink} Spin drag was first considered in semiconductors,\cite{scd_giovanni} where it gives rise to a temperature-dependent difference between spin and charge diffusion constants. This latter difference was indeed observed experimentally. \cite{weber_nature_2005} For ultracold fermions with repulsive interactions, the one-dimensional situation was discussed in Ref.~[\onlinecite{marco}] and an enhancement of the spin-drag resistivity was predicted close to the ferromagnetic (Stoner) transition \cite{duinemag} and Bardeen-Cooper-Schrieffer (BCS) transition. \cite{mink2012} Enhancement of the collision rate was observed in the BCS regime. \cite{riedl} Experimentally, the spin-drag resistivity was measured in the  strongly  interacting (unitary) regime of fermionic cold atoms but an enhancement was not clearly observed.\cite{sommer} For ultracold bosons close to the Bose-Einstein condensation transition, Bose-enhanced scattering between atoms was predicted to lead to an enhanced spin-drag resistivity in three dimensions, \cite{hedwig} which was indeed observed experimentally. \cite{koller2012}

In this work we present a unified theory of drag phenomena near a second-order phase transition -- a theory that encompasses the effects described in the previous paragraph and provides a  framework for the study of similar effects yet to be discovered. This theory is based on quantum kinetic theory and Fermi's golden rule for the scattering amplitudes. Within these approximations, the theory is valid in both two and three dimensions, for particle- and hole-like bands, and for arbitrary imbalances in density and mass. We show how to incorporate the effects of Gaussian critical fluctuations close to a phase transition due to an instability in a specific (Hartree, Fock, or Cooper) channel.
More precisely, adopting the Gaussian model of critical fluctuations
means that the dominant energy and wave vector dependence of the quasiparticle scattering amplitude near the phase transition is calculated in terms of the non-interacting, i.e., Gaussian,  propagator of the fluctuations of the order parameter.   The benefit of this approach is that we
are therefore able to incorporate these fluctuations within the transparent Boltzmann formalism which allows a straightforward calculation of the transport coefficients.
Truly critical fluctuations, resulting from interactions between order-parameter
fluctuations, are not straightforwardly taken into account within this formalism as it relies on a quasi-particle description, which  usually breaks down near the phase transition. When critical fluctuations beyond the Gaussian level are important the diagrammatic approach is a more natural starting point.\cite{rakpong} Such fluctuations are important only in the Ginzburg region,\cite{amitbook} where our Boltzmann approach ceases to be a good approximation.

The purpose of this article is twofold: on the one hand, we present a unified view of the special cases considered in our previous publications.\cite{duinemag,mink,mink2012} On the other hand, we present new details and improved results from our general formalism for three-dimensional Fermi gases and new results for electron-hole bilayers.

The remainder of the paper is organized as follows. In Sec.~\ref{sec:ss} we introduce our formalism in the simple case of the Boltzmann equation for a single species. In Sec.~\ref{sec:BM} we solve the coupled Boltzmann equations of the two species and find an expression for the drag resistivity in terms of the collision integral. We derive a general expression for this collision integral in Sec.~\ref{sec:CI}. Our results for quadratic-dispersion systems, e.g. electrons in semiconductors and cold atoms, are given in Sec.~\ref{sec:quad} and for linear-dispersion systems, e.g. massless Dirac fermions in graphene, in Sec.~\ref{sec:lin}. In Sec.~\ref{sec:allresults} we present analytical results for the behavior of the drag resistivity close to the critical temperature, as well as numerical results for specific systems. Our conclusions are in Sec.~\ref{sec:con}. An appendix is included that details some calculational steps that were skipped in the main text, but may yet benefit the reader who is interested in applying the theory to other systems.

\section{Single species results}
\label{sec:ss}
\begin{figure}
\begin{center}
\includegraphics[width = 0.5 \textwidth]{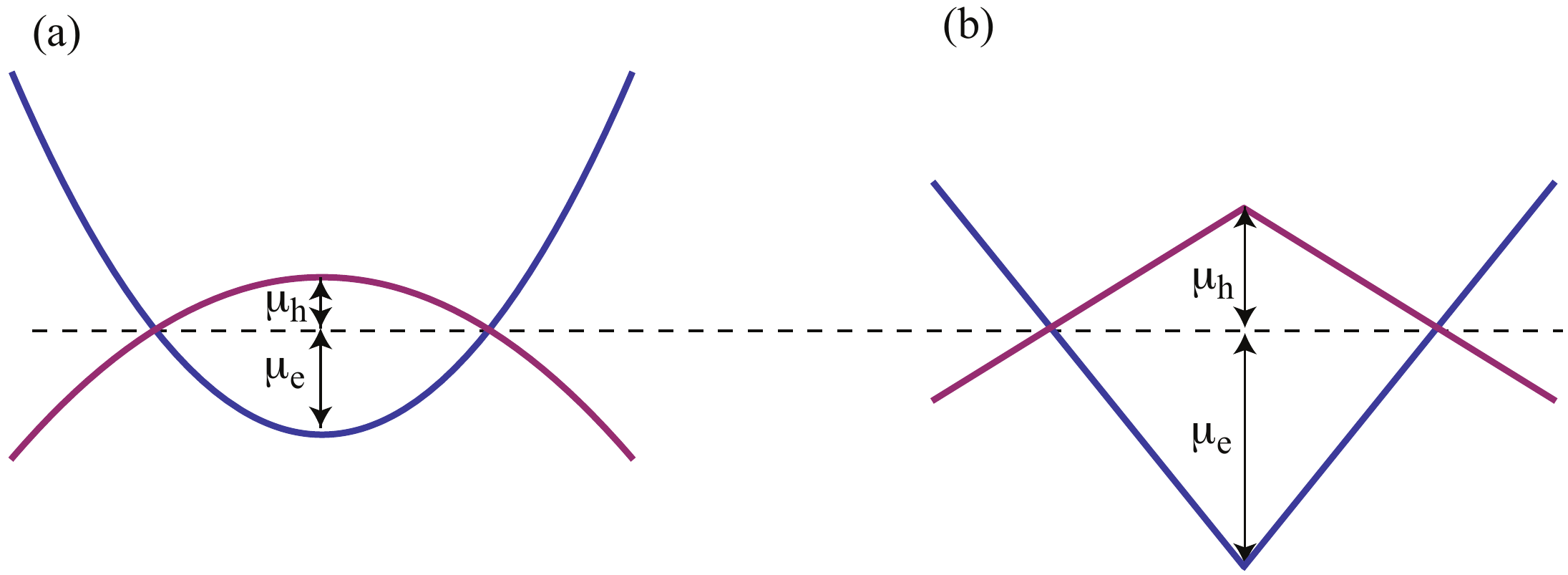}
\caption{ \label{fig:dispersion} Electron and hole bands are drawn in red and blue respectively for parabolic dispersion (a) and linear dispersion (b). The chemical potentials are also indicated: $\mu_{\rm e}$ is the electron chemical potential, whereas $\mu_{\rm h}$ denotes minus the hole chemical potential.}
\end{center}
\end{figure}

 We start by specifying the single-particle dispersion relations that we will use:
\begin{equation} \label{eq:dispS}
\xi({\bm k}) = s (\hbar^2 k^2/2 m - \mu) \quad \text{or} \quad
\xi({\bm k}) = s (\hbar v k - \mu)~,
\end{equation}
for a quadratic and linear dispersion, respectively. Here, $\hbar k$  is the momentum of the carrier and $m$ their mass (in case of a parabolic band). Alternatively, $v$ is the carrier velocity for linear dispersion. When the band is ``particle-like" $s=1$ and when the band is ``hole-like" $s=-1$. Alternatively, we will say that for $s=1$ the band has a positive sign and for $s=-1$ a negative sign. Note that when $s = -1$ the $\mu$ we introduce is actually the negative of the chemical potential measured from the top of the hole band (see Fig.~\ref{fig:dispersion}).

Before writing two coupled Boltzmann equations to calculate the drag resistivity, it is instructive to consider the ordinary resistivity of a single species. Along the way we define some quantities we need later on. To determine this resistivity, we need to determine the non-equilibrium carrier distribution $f$ in the presence of a uniform force field  ${\bm F}$. Note that, since we consider both mass transport (in the case of cold-atom systems) and charge  transport (in the case of solid-state systems), we prefer to keep the discussion general and use force rather than electric field ${\bm E}$. In the case of electrons the force is of course equal to ${\bm F} = -|e| {\bm E}$, with $-|e|$ the charge of a single electron. The distribution function $f({\bm k}(t))$ is independent of position and, in the absence of relaxation, obeys the equation of motion
\be
\frac{d}{d t}  f({\bm k}(t)) =  \dot{{\bm k}} \cdot  \partial_{\bm k} f({\bm k})~.
\ee
In the relaxation-time approximation, one adds a phenomenological term which relaxes $f$ back to the equilibrium Fermi-Dirac distribution on a time scale $\tau$. As we are interested in drag effects due to interactions we take the simplest version of the relaxation-time approximation and ignore momentum dependence of the relaxation time. Under the influence of a force ${\bm F}$, we have $\hbar \dot{{\bm k}} = {\bm F}$. Then, for a steady-state solution
\begin{equation} \label{eq:slBM}
\frac{1}{\hbar} {\bm F} \cdot  \partial_{\bm k} f({\bm k}) = -\frac{1}{\tau}  [f({\bm k}) - n_{\rm F}(\xi({\bm k}))]~,
\end{equation}
where $\xi({\bm k})$ is the bare dispersion introduced above and $n_{\rm F}(\epsilon) = 1/(1+\exp( \beta \epsilon))$ is the Fermi-Dirac distribution with $\beta = (k_{\rm B} T)^{-1}$ the inverse thermal energy.

We are interested in linear-response transport coefficients, so we take as an {\it ansatz} for the solution of Eq.~(\ref{eq:slBM}) the first-order expansion of a Fermi-Dirac distribution shifted by a drift momentum $\hbar {\bm k}^\text{drift}$:
\ber\label{eq:f1}
f({\bm k})
&= &n_{\rm F}[\xi({\bm k})] -  s [{\bm k}^\text{drift} \cdot \partial_{\bm k} \xi({\bm k})]~n_{\rm F}'[\xi({\bm k})] \nonumber\\
& \equiv& n_{\rm F}(\xi({\bm k})) + f^{(1)}({\bm k})~,
\eer
where $n_{\rm F}'(\ep) = \partial_\ep n_{\rm F}(\ep)$.
The inclusion of the extra factor $s$ ensures that the average momentum of the ensemble of particles is proportional to the carrier density $n$, i.e., the number of electrons when $s=1$ and the number of holes when $s = -1$, so that
\be
  \frac{1}{V} \sum_{{\bm k}} \hbar {\bm k} f^{(1)}({\bm k}) =  \hbar {\bm k}^\text{drift} n~.
\ee

We also need to evaluate the current density ${\bm j}$:
\be
{\bm j} = \frac{1}{V \hbar} \sum_{{\bm k}} [\partial_{\bm k} \xi({\bm k})] f^{(1)}({\bm k})~.
\ee
Note that, for the same reasons as mentioned before we consider mass, rather than charge current, and omit a prefactor $-|e|$. As a consequence, for the case of electrons the resistivities found below should be multiplied with a factor $e^2$ to convert them to electrical resistivities. Introducing the current-to-momentum conversion factor $C$
\be
C =  -  \frac{1}{d V \hbar} \sum_{{\bm k}} [\partial_{\bm k} \xi({\bm k})]^2~n_{\rm F}'(\xi_{s}({\bm k}))~,
\ee
where $d$ is the dimensionality, we obtain ${\bm j} = s C {\bm k}^\text{drift}$. The conversion factor is easily determined to be
\be
C =  \frac{\hbar n}{m} \quad \text{or} \quad
C =  \frac{\mu}{4 \pi \hbar}~,
\ee
for a quadratic and linear dispersion, respectively.

Using these definitions, we solve the Boltzmann equation Eq.~(\ref{eq:slBM}) which, in the linear-response approximation, reads
\be\label{eq:slBMfo}
\frac{1}{\hbar} {\bm F} \cdot  \partial_{\bm k} n_{\rm F}(\xi({\bm k})) = -\frac{1}{\tau}  f^{(1)}({\bm k})~.
\ee
Performing the differentiation with respect to ${\bm k}$ on the left-hand side of the above leads to ${\bm k}^\text{drift}   =s (\tau/\hbar) {\bm F}$ and ${\bm F} = (\hbar/C \tau) {\bm j}$, from which we identify the resistivity ${\bm F} = \rho {\bm j}$ as
\be
\rho = \frac{m}{ n \tau}
\quad \text{and} \quad
\rho = \frac{4 \pi \hbar^2}{\mu \tau}~,
\ee
for a quadratic and linear dispersion, respectively. For the quadratic dispersion we recognize the familiar Drude result.

\section{Drag resistivity and coupled Boltzmann equations}
\label{sec:BM}

In this section we first introduce and then solve two coupled Boltzmann equations and determine the drag resistivity which is the focus of this article. We denote the two species by the pseudospin label $\sigma = \uparrow,\downarrow$ which can either be hyperfine spin for the case of cold Fermi gases or layer index in the case of double-layer systems. The dispersions in Eq.~(\ref{eq:dispS}) acquire the species label $\sigma$ and are
\begin{equation} \label{eq:disp}
\xi_\sigma ({\bm k}) = s_\sigma (\hbar^2 k^2/2 m_\sigma - \mu_\sigma) \quad \text{or} \quad
\xi_\sigma ({\bm k}) = s_\sigma (\hbar v k - \mu_\sigma)~,
\end{equation}
for the quadratic and linear dispersion, respectively. Note that we allow for a mass ``imbalance" (i.e. $m_\uparrow \neq m_\downarrow$) and population imbalance (i.e. $\mu_\uparrow \neq \mu_\downarrow$).

We will denote the number of degenerate fermion types in a species by $N_f$, which will always be equal for both species. For example, in an electron-hole double-layer, $N_f = 2$, because both layers have spin degeneracy. In double-layer graphene (two graphene sheets separated by a tunnel barrier), $N_f = 4$, due to the presence of spin degeneracy and two Dirac cones in each layer. The density $n$ is always the density of a single fermion type, the total density of species $\sigma$ is $N_f n_\sigma$ and the total carrier density in the system $N_f(n_\uparrow + n_\downarrow)$. We apply a species-dependent force ${\bm F}_\sigma$ so that the equivalent of the linearized Boltzmann equation in Eq.~(\ref{eq:slBMfo}) is the following system of two coupled equations:
\begin{align}
\frac{1}{\hbar} {\bm F}_\uparrow   \cdot \partial_{\bm k} n_{\rm F}(\xi_\uparrow({\bm k}))   &=- \frac{1}{\tau_\uparrow} f^{(1)}_\uparrow({\bm k}) + \Gamma_\uparrow({\bm k})~;\\
\frac{1}{\hbar} {\bm F}_\downarrow \cdot \partial_{\bm k} n_{\rm F}(\xi_\downarrow({\bm k})) &=- \frac{1}{\tau_\downarrow} f^{(1)}_\downarrow({\bm k}) + \Gamma_\downarrow({\bm k})~,
\end{align}
where we introduced species-dependent intra-species relaxation times $\tau_\sigma$. The $\Gamma_\sigma({\bm k})$ are the collision integrals which give the net flux of particles into the state ${\bm k}$ of species $\sigma$. We will specify them in the next section. We substitute the expressions for $f^{(1)}_\sigma$ to introduce the drift momenta ${\bm k}^\text{drift}_\sigma$
\begin{align}
\left[\frac{1}{\hbar} {\bm F}_\uparrow  - \frac{s_\uparrow}{\tau_\uparrow} {\bm k}^\text{drift}_\uparrow \right]\cdot [\partial_{\bm k} \xi_\uparrow({\bm k})] n_{\rm F}'(\xi_\uparrow({\bm k})) &=  \Gamma_\uparrow({\bm k})~;\\
\left[\frac{1}{\hbar} {\bm F}_\downarrow  - \frac{s_\downarrow}{\tau_\downarrow} {\bm k}^\text{drift}_\downarrow \right]\cdot [\partial_{\bm k} \xi_\downarrow({\bm k})] n_{\rm F}'(\xi_\downarrow({\bm k})) &=  \Gamma_\downarrow({\bm k})~.
\end{align}
To make connection to the current density, we multiply with the group velocity and sum over ${\bm k}$ with the result
\begin{align}
-  C_{\uparrow} \left[\frac{1}{\hbar} {\bm F}_\uparrow  - \frac{s_\uparrow}{\tau_\uparrow} {\bm k}^\text{drift}_\uparrow \right] &
=  {\bm \Gamma}_\uparrow~;\\
-  C_{\downarrow}\left[\frac{1}{\hbar} {\bm F}_\downarrow  - \frac{s_\downarrow}{\tau_\downarrow} {\bm k}^\text{drift}_\downarrow \right] &
=  {\bm \Gamma}_\downarrow~,
\end{align}
where we defined ${\bm \Gamma}_\sigma = (N_f^2 /V\hbar)\sum_{\bm k} (\partial_{\bm k} \xi_\sigma({\bm k})) \Gamma_\sigma ({\bm k})$.

Below, we find that to first order in the drift momenta the above yields
${\bm \Gamma}_\uparrow  = N_f^2(s_\uparrow \Gamma^{\rm S}_\uparrow {\bm k}^\text{drift}_\uparrow + s_\downarrow \Gamma^{\rm D} {\bm k}^{\rm drift}
_\downarrow)$ and ${\bm \Gamma}_\downarrow  = N_f^2(s_\downarrow \Gamma^{\rm S}_\downarrow {\bm k}^{\rm drift}
_\downarrow + s_\uparrow \Gamma^{\rm D} {\bm k}^{\rm drift}_\uparrow)$,
where ``S" labels the contribution of collisions between particle of the same species, and ``D"   labels the contribution of collisions from particles of different species.   Note that the coefficients for the cross dependence are equal in both relations (i.e., $\Gamma^{\rm D}$ does not depend on pseudospin).
After substituting these expansions, we obtain the resistivity matrix relating ${\bm F}$ to ${\bm j}$
\be
\begin{pmatrix} {\bm F}_\uparrow \\ {\bm F}_\downarrow \end{pmatrix} = \begin{pmatrix}
\rho_{\uparrow\uparrow} & \rho_{\uparrow\downarrow} \\
\rho_{\downarrow\uparrow} & \rho_{\downarrow\downarrow}
\end{pmatrix} \begin{pmatrix} {\bm j}_\uparrow \\ {\bm j}_\downarrow \end{pmatrix},
\ee
with the drag resistivity $\rho_{\rm D} \equiv \rho_{\downarrow\uparrow} = \rho_{\uparrow\downarrow} = - N_f \hbar \Gamma^{\rm D}/ C_{\downarrow} C_{\uparrow}$. The intra-species resistivities are $\rho_{\sigma \sigma} = \hbar/N_f C_\sigma \tau_\sigma - N_f \hbar \Gamma^{\rm S}_\sigma/C^2_\sigma$. When the interspecies collision integrals are zero, $\rho_{\sigma \sigma}$ reduces to the result obtained in the previous section for the single-species problem and $N_f=1$.

\section{Collision Integral}
\label{sec:CI}
In this section we determine the collision integral ${ \bm \Gamma}_\sigma$. We start with the expression for the scattering rate from Fermi's golden rule and expand it to first order in the drift momenta. Let $\uparrow,{\bm k}_1$ and $\downarrow,{\bm k}_2$ be the incoming states, which are scattered onto the final states $\downarrow,{\bm k}_3$ and $\uparrow,{\bm k}_4$. The rate for this process can be calculated from Fermi's golden rule:
\ber \label{eq:rate}
R &=& \frac{2 \pi N_f^2}{\hbar V^2} |W({\bm k}_1,{\bm k}_2,{\bm k}_3,{\bm k}_4)|^2 \delta_{{\bm k}_1 + {\bm k}_2,{\bm k}_3 + {\bm k}_4}\nonumber \\
&\times&\delta(\xi_\uparrow({\bm k}_1) + \xi_\downarrow({\bm k}_2) - \xi_\downarrow({\bm k}_3) - \xi_\uparrow({\bm k}_4)) \nonumber\\
&\times& f_\uparrow({\bm k}_1) f_\downarrow({\bm k}_2) (1-f_\downarrow({\bm k}_3 )) (1- f_\uparrow({\bm k}_4))~.
\eer
We note the presence of the standard Fermi's golden rule factors in Eq.~(\ref{eq:rate}): the delta functions ensuring energy and momentum conservation and the interaction matrix element $W$ squared. Additionally, we see that each contribution is weighted with the distribution functions of the two incoming particles, and with one minus the distribution functions for the two outgoing particles. For two-dimensional chiral electron systems described by a massless Dirac equation, such as electrons in double-layer graphene or topological insulator thin films, we should multiply the right-hand side of the last equation with the following form factors
\be \label{eq:ff}
\frac{1+\cos(\phi_4 - \phi_1)}{2} \frac{1+\cos(\phi_3 - \phi_2)}{2}~,
\ee
with $\phi_i$ the angle between ${\bm k}_i$ and the $x$-axis. These form factors encode suppressed backscattering in chiral electron systems. For simplicity, we will drop these factors in the manipulations carried out in this section, and reinstate them when we consider systems with linear dispersions in section~\ref{sec:lin}.

To make our manipulations more tractable, we introduce some shorthand notations: for $i=1,4$ we have $\xi_i = \xi_\uparrow({\bm k}_i)$ and $f_i = f_\uparrow({\bm k}_i)$, and for $i=2,3$ we have $\xi_i = \xi_\downarrow({\bm k}_i)$ and $f_i = f_\downarrow({\bm k}_i)$. Furthermore, the delta-function expressing the conservation of momentum ($\delta_{{\bm k}_1 + {\bm k}_2,{\bm k}_3 + {\bm k}_4}$) will be indicated by
$\delta_{{\bm k}_i}$, while the one expressing the conservation of energy [$\delta(\xi_\uparrow({\bm k}_1) + \xi_\downarrow({\bm k}_2) - \xi_\downarrow({\bm k}_3) - \xi_\downarrow({\bm k}_4))$] by $\delta(\xi_i)$. Using this new notation and dropping the arguments of the many-particle scattering amplitude $W$, the rate $R$ becomes
\be
R= \frac{2 \pi N_f^2}{\hbar V^2} \delta_{{\bm k}_i}\delta(\xi_i) |W|^2 f_1 f_2 (1-f_3) (1- f_4).
\ee
For concreteness, we will now determine the collision integral for the $\uparrow$-species. The expression for the $\downarrow$-species can be easily obtained by appropriately changing  the labels. The net flux of particle into the state $\uparrow,{\bm k}$ which is $\Gamma_\uparrow({\bm k})$ is then
\be
\Gamma_\uparrow({\bm k}) = \sum_{{\bm k}_1,{\bm k}_2,{\bm k}_3,{\bm k}_4} R (\delta_{{\bm k}_4,{\bm k}} - \delta_{{\bm k}_1,{\bm k}}).
\ee
The expression for ${\bm \Gamma}_\uparrow$ is then
\ber \label{eq:gav}
{\bm \Gamma}_\uparrow &=& \frac{N_f^2}{V\hbar}\sum_{\bm k} (\partial_{\bm k} \xi_\uparrow({\bm k})) \Gamma_\uparrow ({\bm k}) =
\frac{2 \pi}{\hbar^2 V^3} \sum_{{\bm k}_i}  \delta_{{\bm k}_i} \delta(\xi_i)\nonumber\\
&\times& |W|^2 f_1 f_2 (1-f_3) (1- f_4)  (\partial_{\bm k} \xi_{4} -\partial_{\bm k} \xi_{1}),
\eer
where $\sum_{{\bm k}_i}$ is a shorthand notation for $\sum_{{\bm k}_1,{\bm k}_2,{\bm k}_3,{\bm k}_4} $ and where $\partial_{\bm k} \xi_{i} \equiv  [\partial_{\bm k} \xi_\uparrow({\bm k})]_{{\bm k}={\bm k}_i}$ and $\partial_{\bm k} \xi_{i} \equiv  [\partial_{\bm k} \xi_\downarrow({\bm k})]_{{\bm k}={\bm k}_i}$ for $i=1,4$ and $i=2,3$, respectively.

We perform the first-order expansion in the drift momenta ${\bm k}^\text{drift}_\sigma$ of the distribution functions $f_\sigma$ and obtain to first order in the $f^{(1)}_\sigma$'s
\ber\label{eq:linearization}
&& f_1 f_2 (1-f_3) (1- f_4) \to \nonumber \\
&& n_1 n_2 (1-n_3 )(1- n_4) \nonumber \\
&\times&\left(\frac{f^{(1)}_1}{n_1} + \frac{f^{(1)}_2}{n_2}  - \frac{f^{(1)}_3}{1-n_3} - \frac{f^{(1)}_4}{1-n_4}\right),
\eer
where we dropped the zeroth-order term since it evaluates to zero in the momentum summation. In the previous equation we introduced the shorthand notation $n_i = n_{\rm F}(\xi_i)$. Substituting Eq.~(\ref{eq:linearization}) and the expression for $f^{(1)}_\sigma$ from Eq.~(\ref{eq:f1}) in Eq.~(\ref{eq:gav}) and using that $n_{\rm F}'(\epsilon) = - \beta n_{\rm F}(\epsilon)[1-n_{\rm F}(\epsilon)]$ yields
\begin{multline}
{\bm \Gamma}_\uparrow  = \frac{2 \pi \beta N_f^2}{\hbar^2 V^3} \sum_{{\bm k}_i}  \delta_{{\bm k}_i} \delta(\xi_i)
|W|^2 n_1 n_2 (1-n_3) (1- n_4)  \\  (\partial_{\bm k} \xi_{4} -\partial_{\bm k} \xi_{1})
\left\{
s_\uparrow {\bm k}^\text{drift}_\uparrow \cdot  \left[(1-n_1)\partial_{\bm k} \xi_{1}  - n_4\partial_{\bm k} \xi_{4}\right] +\right.\\
\left. s_\downarrow {\bm k}^\text{drift}_\downarrow \cdot \left[(1-n_2) \partial_{\bm k} \xi_{2} -  n_3\partial_{\bm k} \xi_{3}\right]
\right\}~.
\end{multline}
When we interchange the labels of the incoming and outgoing states (specifically, when we transform ${\bm k}_1 \leftrightarrow {\bm k}_4$ and ${\bm k}_2 \leftrightarrow {\bm k}_3$), ${\bm \Gamma}_\uparrow$ remains invariant. Taking the average of the original expression for ${\bm \Gamma}_\uparrow $ and the one transformed in this way, and using that due to energy conservation $n_1 n_2 (1-n_3) (1- n_4)  = n_4 n_3 (1-n_2) (1- n_1)$, and also that the interaction $W$ remains invariant under this transformation, we find that
\ber\label{eq:gav2}
{\bm \Gamma}_\uparrow  &=& - \frac{\pi \beta N_f^2}{\hbar^2 V^3} \sum_{{\bm k}_i}  \delta_{{\bm k}_i} \delta(\xi_i)
|W|^2 n_1 n_2 (1-n_3) (1- n_4)  \nonumber \\
&\times& (\partial_{\bm k} \xi_{4} -\partial_{\bm k} \xi_{1})
\big[s_\uparrow {\bm k}^{\rm drift}_\uparrow \cdot (\partial_{\bm k} \xi_{4}  - \partial_{\bm k} \xi_{1}) \nonumber\\
&+& s_\downarrow {\bm k}^{\rm drift}_\downarrow \cdot (\partial_{\bm k} \xi_{3} - \partial_{\bm k} \xi_{2})\big]~.
\eer
In the Appendix we show that we may take the drift momenta out of the summation provided that we introduce a factor $1/d$, so that we obtain
\begin{align}
{\bm \Gamma}_\uparrow  &= N_f^2 (s_\uparrow \Gamma^{\rm S}_\uparrow {\bm k}^{\rm drift}_\uparrow
+ s_\downarrow \Gamma^{\rm D} {\bm k}^\text{drift}_\downarrow)~; \\
{\bm \Gamma}_\downarrow &= N_f^2 (s_\downarrow \Gamma^{\rm S}_\downarrow {\bm k}^{\rm drift}_\downarrow + s_\uparrow \Gamma^{\rm D} {\bm k}^\text{drift}_\uparrow)~,
\end{align}
where we again note that the cross contributions have the same coefficient $\Gamma^{\rm D}$. The explicit expressions for these coefficients are
\begin{multline} \label{eq:gasu}
\Gamma^{\rm S}_\uparrow  = -\frac{ \pi \beta}{ d \hbar^2 V^3} \sum_{{\bm k}_i}  \delta_{{\bm k}_i} \delta(\xi_i)
|W|^2 n_1 n_2 (1-n_3) (1- n_4) \\
(\partial_{\bm k} \xi_{4} -\partial_{\bm k} \xi_{1})^2~,
\end{multline}
\begin{multline} \label{eq:gasd}
\Gamma^{\rm S}_\downarrow = - \frac{ \pi \beta}{ d \hbar^2 V^3} \sum_{{\bm k}_i}  \delta_{{\bm k}_i} \delta(\xi_i)
|W|^2 n_1 n_2 (1-n_3) (1- n_4) \\(\partial_{\bm k} \xi_{3} -\partial_{\bm k} \xi_{2})^2~,
\end{multline}
and
\begin{multline} \label{eq:gad}
\Gamma^{\rm D} = - \frac{ \pi \beta}{ d \hbar^2 V^3} \sum_{{\bm k}_i}  \delta_{{\bm k}_i} \delta(\xi_i)
|W|^2 n_1 n_2 (1-n_3) (1- n_4) \\ (\partial_{\bm k} \xi_{4} -\partial_{\bm k} \xi_{1})  \cdot (\partial_{\bm k} \xi_{3} - \partial_{\bm k} \xi_{2})~.
\end{multline}
We note that these results and the expression for the drag resistivity $\rho_{\rm D} = - \hbar \Gamma^{\rm D}/  C_{\downarrow} C_{\uparrow}$ are valid for two and three dimensions, for a linear (after reinstating  the form factors Eq.~(\ref{eq:ff}) in needed) and quadratic dispersion, for particle-like  and hole-like bands, and arbitrary density and mass imbalance.

We now comment on the signs of $\Gamma^{\rm D}$ and $\rho_{\rm D}$. We stress that $\rho_{\rm D}$ is the transport coefficient relating the mass (and not the charge) current to the force. The sign of $\Gamma^{\rm D}$ is $s_\uparrow s_\downarrow$, as can be seen by considering a quadratic dispersion, so that the sign of $\rho_{\rm D}$ is $- s_\uparrow s_\downarrow$. Consider the situation in which there is a non-zero current in the ``active" $\uparrow$ species ${\bm j}_\uparrow$, and that the current in the passive $\downarrow$ species is held to zero by the force ${\bm F}_\downarrow$, so that ${\bm F}_\downarrow =  \rho_{\rm D} {\bm j}_\uparrow$. Recall that the intra-species resistivities are positive for both particle-like and hole-like bands. For bands of equal character, interaction tend to equalize the currents in both layers (or spin projections). Thus, for bands of equal character, $\rho_{\rm D}$ is negative. When one of the bands is particle-like and the other hole-like, $\rho_{\rm D}$ is positive.

It is interesting to consider the relations between $\Gamma^{\rm S}_{\uparrow,\downarrow}$ and $\Gamma^{\rm D}$. For a quadratic dispersion we substitute Eq.~(\ref{eq:disp}) into Eqs.~(\ref{eq:gasu},\ref{eq:gasd},\ref{eq:gad}) and obtain
\begin{align}
{\bm \Gamma}_\uparrow    &= -s_\downarrow m_\downarrow N_f^2|\Gamma^{\rm D}|  \left( \frac{{\bm k}^\text{drift}_\uparrow}{m_\uparrow} -   \frac{{\bm k}^\text{drift}_\downarrow}{m_\downarrow} \right)  \\
{\bm \Gamma}_\downarrow  &= -s_\uparrow m_\uparrow N_f^2|\Gamma^{\rm D}|  \left(\frac{{\bm k}^\text{drift}_\downarrow}{m_\downarrow} -  \frac{{\bm k}^\text{drift}_\uparrow}{m_\uparrow}\right)~.
\end{align}
When, ${\bm k}^\text{drift}_\uparrow/m_\uparrow = {\bm k}^\text{drift}_\downarrow/m_\downarrow$ the integrated collision integral vanishes and the resistivities are just given by the decoupled single-species result. This condition is satisfied when the drift velocities have the same magnitude, and the drift momenta have the same direction. This is what one expects for a Galilean invariant system.
For the case of a linear dispersion the Hamiltonian is not Galilean invariant and no simple relations between the $\Gamma^{\rm S}_\sigma$
and $\Gamma^{\rm D}$ has been obtained.

\subsection{Instability channels}
At the methodological level, the main purpose of this article is to show how to determine the collision integral $\Gamma^{\rm D}$ in Eq.~(\ref{eq:gad}) incorporating the effect of Gaussian fluctuations close to a phase transition. In the case of a ferromagnetic transition these are magnetic fluctuations, while in the case of superconductivity or exciton condensation, these are pairing fluctuations. These fluctuations increase in strength when the system approaches the transition and ultimately lead to a divergence of the scattering amplitude at the critical temperature $T_{\rm c}$. This divergence occurs at energy $\hbar \omega=0$ and momentum ${\bm k}_{\rm p}=0$, that are related to the energies and momenta of the incoming particles in a way that depends on the channel in which the instability occurs. To take into account these fluctuations and account for the dependence of the scattering amplitude on $\hbar \omega$ and ${\bm k}_{\rm p}$ that is dominant close to the phase transition, we should ``decouple" Eq.~(\ref{eq:gad}) in the correct channel. This decoupling is carried out by introducing the appropriate auxiliary energy variable $\hbar\omega$ through an integral over it. For example, in the case of superconductivity between two bands with the same character, the pole in the scattering amplitude (in this case it is usually called the ``many-body $T$ matrix") lies at zero energy and zero center-of-mass momentum of the two incoming particles $\hbar\omega = \xi_1 + \xi_2 = 0$ and ${\bm k}_{\rm p} ={\bm k}_1 + {\bm k}_2= {\bm 0}$. Thus, in this specific case, we would introduce $\omega$ as follows
\ber
\delta(\xi_1 + \xi_2 - \xi_3 - \xi_4) &=& \int d(\hbar \omega)~\delta(\xi_1 + \xi_2 - \hbar \omega) \nonumber\\
&\times &\delta(\xi_3 + \xi_4 - \hbar \omega)~.
\eer

\begin{figure}
\begin{center}
\includegraphics[width = 0.3 \textwidth]{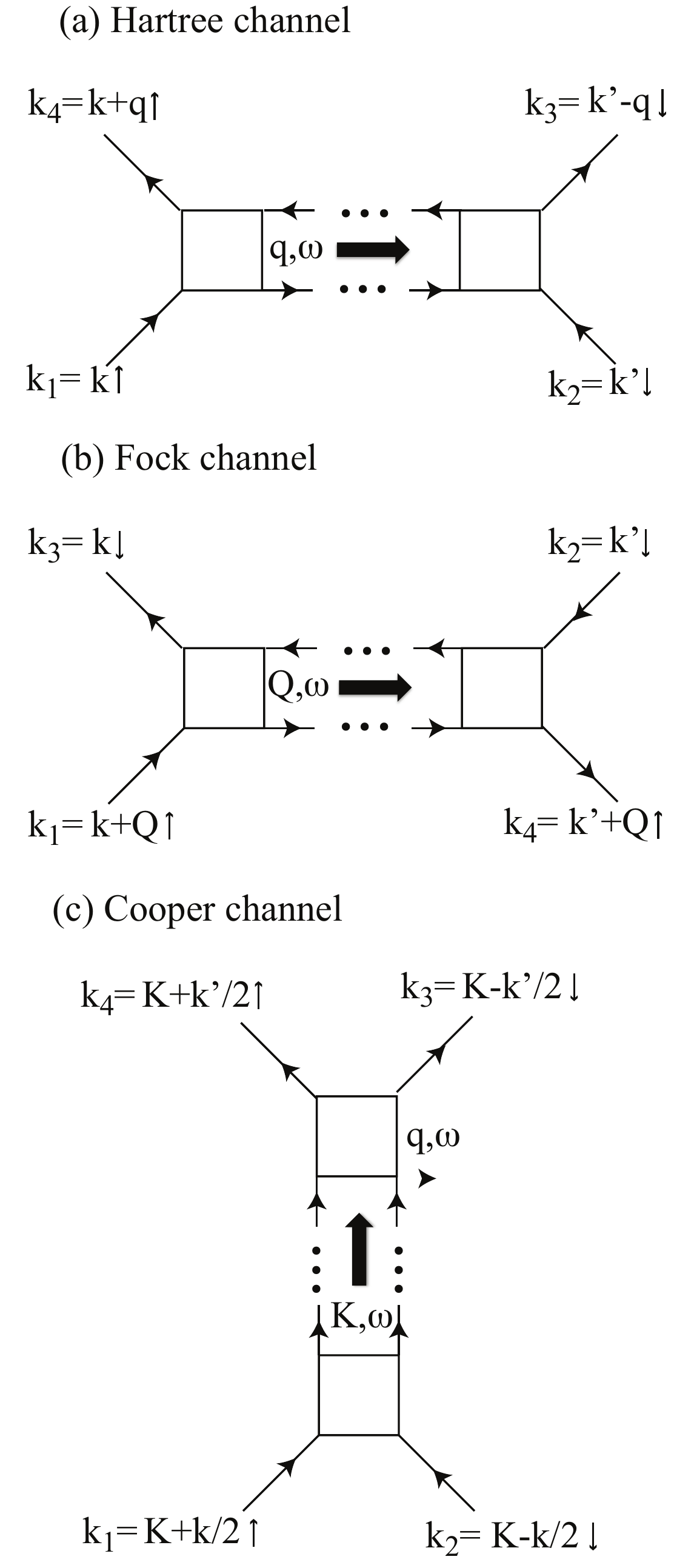}
\caption{\label{fig:Channels}The three channels, Hartree (a), Fock (b), Cooper (c) in which the scattering amplitudes can diverge for $q,\omega \to 0$, $Q,\omega \to 0$ and  $K,\omega \to0$ respectively.  Notice that the physical momentum transferred in the scattering process, $k_4-k_1$, coincides with the channel momentum, $q$,  in the Hartree channel, but not in the Fock and Cooper channels.}
\end{center}
\end{figure}

The three possibilities of combining the incoming $\uparrow$ particle with energy $\xi_1$ with i) the incoming $\downarrow$ particle with energy $\xi_2$, ii) the outgoing $\downarrow$ particle with energy $\xi_3$, and iii) the outgoing $\uparrow$ particle with energy $\xi_4$, are denoted as decoupling in the Cooper, Fock, and Hartree channels, respectively. We summarize the properties of these instability channels in Table~\ref{tab:1} and give their Feynman diagrams in Fig.~\ref{fig:Channels}.

\begin{table}
\begin{center}
\begin{tabular}{l|l|c|c}
Instability channel     & Combination                           &$\hbar \omega$        & ${\bm k}_p$                                \\ \hline
Hartree                 & In $\uparrow$ with Out $\uparrow$     &$\xi_1 - \xi_4$       & ${\bm k}_1 - {\bm k}_4 \equiv {\bm q}$     \\
Fock                    & In $\uparrow$ with Out $\downarrow$   &$\xi_1 - \xi_3$       & ${\bm k}_1 - {\bm k}_3 \equiv {\bm Q}$     \\
Cooper                  & In $\uparrow$ with In $\downarrow$    &$\xi_1 + \xi_2$       & ${\bm k}_1 + {\bm k}_2 \equiv {\bm K}$
\end{tabular}
\end{center}
\caption{\label{tab:1} Overview of the three instability channels. In the third and fourth columns we specify the energy and momentum variables appropriate for this instability channel.  }
\end{table}

In the next section, we discuss how to perform the decoupling of Eq.~(\ref{eq:gad}) in the three channels in full detail for a quadratic dispersion. These results are valid for arbitrary spatial dimensionality, band character, mass, and density imbalance. In section~\ref{sec:lin} we consider a linear dispersion and restrict ourselves to the evaluation of Eq.~(\ref{eq:gad}) in a system with exciton-condensation, i.e., the decoupling of Eq.~(\ref{eq:gad}) in the Fock channel for bands of opposite character. Then, in section~\ref{sec:res} we give some qualitative results and present quantitative results for the resistivity of three-dimensional cold Fermi gases close to a Stoner transition and two-dimensional spatially-separated electron and hole systems in semiconductor double quantum wells. A list of phase transitions in various systems and their instability channels can be found in Table~\ref{tab:2}.

\begin{table}
\begin{center}
\begin{tabular}{l|l|l|l}
System                      & Transition                & $s_\uparrow,s_\downarrow$     & Channel \\ \hline
Cold gases                  & Magnetism longitudinal    & 1,1                           & Hartree \\
Cold gases                  & Magnetism transverse      & 1,1                           & Fock \\
Cold gases                 & BCS                       & 1,1                           & Cooper \\
Electron-hole bilayers       & Exciton condensation      & 1,-1                          & Fock \\
TI thin films                & Exciton condensation      & 1,-1                          & Fock
\end{tabular}
\end{center}
\caption{\label{tab:2} Overview of phase transitions in various systems and their instability channels.}
\end{table}

\section{Quadratic Dispersion}
\label{sec:quad}

For a quadratic dispersion, Eq.~(\ref{eq:gad}) simplifies to
\begin{multline} \label{eq:gadq}
\Gamma^{\rm D} = s_\uparrow s_\downarrow \frac{\pi \beta}{d V^3} \frac{\hbar^2}{m_\uparrow m_\downarrow} \sum_{{\bm k}_i}  \delta_{{\bm k}_i} \delta(\xi_i)\\
|W|^2 n_1 n_2 (1-n_3) (1- n_4) ({\bm k}_1 - {\bm k}_4)^2~,
\end{multline}
where we note that ${\bm k}_1 - {\bm k}_4$ is just the momentum transferred in the scattering event. Before we discuss the various instabilities in different channels we note that when we replace the scattering amplitude $W$ by a constant, all three instability channels should give the same result for $\Gamma^{\rm D}$. This serves as an important check for the, sometimes lengthy, analytical calculations presented below.

\subsection{Hartree instability}
\label{sec:qh}
When the dominant energy and momentum dependence of the effective interaction $W$ is on the transferred energy $\hbar \omega = \xi_1 - \xi_4$ and momentum ${\bm k}_{\rm p} = {\bm k}_1 - {\bm k}_4 \equiv {\bm q}$ in the scattering event, the Hartree channel is the appropriate instability channel. We introduce an additional integral over $\hbar \omega$ using the identity
\ber
\delta(\xi_1 + \xi_2 - \xi_3 - \xi_4) &=& \int d(\hbar \omega)~\delta(\xi_1 - \xi_4 - \hbar \omega) \nonumber\\
&\times &\delta(\xi_2 - \xi_3 + \hbar \omega)~,
\eer
and solve the delta function for momentum conservation by writing ${\bm k}_1 = {\bm k}$, ${\bm k}_2 = {\bm k}'$, ${\bm k}_3 = {\bm k}'+{\bm q}$, and ${\bm k}_4={\bm k}-{\bm q}$, so that the ${\bm k}$ and ${\bm k}'$ summations in Eq.~(\ref{eq:gadq}) decouple:
\ber
\Gamma^{\rm D} &=& s_\uparrow s_\downarrow  \frac{\pi \beta \hbar^2}{d V m_\uparrow m_\downarrow}  \sum_{{\bm q}} \int d(\hbar \omega)~q^2~|W({\bm q},\hbar \omega)|^2\nonumber \\
&\times & \frac{1}{V} \sum_{{\bm k}} \delta(\xi_\uparrow({\bm k}) - \xi_\uparrow({\bm k}-{\bm q}) - \hbar \omega)
n_{\rm F}(\xi_\uparrow({\bm k}))\nonumber\\
&\times& [1- n_{\rm F}(\xi_\uparrow({\bm k}-{\bm q}))] \nonumber \\
&\times &\frac{1}{V} \sum_{{\bm k}'} \delta(\xi_\downarrow({\bm k}') - \xi_\downarrow({\bm k}'+{\bm q}) + \hbar \omega)
n_{\rm F}(\xi_\downarrow({\bm k}')) \nonumber \\
&\times&[1-n_{\rm F}(\xi_\downarrow({\bm k}'+{\bm q}))]~.
\eer
Note that in the above result we have now explicitly indicated the dominant dependence of the scattering amplitude on ${\bm q}$ and $\hbar \omega$ and ignored any other dependence. We will do this also for the other channels discussed below.  Making use of the identities
$$
n_{\rm F} (x)[1-n_{\rm F} (y)]=[n_{\rm F} (y)-n_{\rm F}(x)]n_{\rm B}(x-y)~,
$$
$$
n_{\rm B}(x)n_{\rm B} (-x)=\frac{1}{4\sinh^2(\beta x/2)}~,
$$
where $n_{\rm B} (x)=[e^{\beta x}-1]^{-1}$ is the Bose distribution, we rewrite this expression as
\begin{multline} \label{eq:gadh}
\Gamma^{\rm D}  = s_\uparrow s_\downarrow  \frac{\beta \hbar^2}{4 d \pi V m_\uparrow m_\downarrow}  \sum_{{\bm q}} \int d \hbar \omega  \frac{q^2  |W({\bm q},\hbar \omega)|^2}{\sinh^2(\beta \hbar \omega/2)}  \\ \Im m\Pi_\uparrow({\bm q},\hbar \omega) \Im m\Pi_\downarrow({\bm q},\hbar \omega)~,
\end{multline}
where we have introduced the well-known polarizability (Lindhard function)
\be\label{eq:PI}
\Pi_\sigma({\bm q},\hbar \omega) =  \frac{1}{V} \sum_{{\bm k}} \frac{n_{\rm F}(\xi_\sigma({\bm k}+{\bm q})) - n_{\rm F}(\xi_\sigma({\bm k}))}{ \xi_\sigma({\bm k}+{\bm q}) - \xi_\sigma({\bm k})-\hbar \omega - i 0^+},
\ee
and we used that $\Pi_\sigma({\bm q},\hbar \omega) = \Pi_\sigma( - {\bm q},\hbar \omega)$ and $\Pi_\sigma({\bm q},-\hbar \omega) = \Pi^*_\sigma({\bm q},\hbar \omega)$.

In the three-dimensional case we have  \cite{giovannibook}
\begin{multline}
\Im m \Pi_\sigma({\bm q},\hbar \omega) = -\frac{m_\sigma^2}{4 \pi \hbar^4 \beta q}
\left\{ \beta \hbar \omega  \right. + \\
\log\left[1 + \exp\left(\beta(\epsilon_\sigma(q/2) + m_\sigma \omega^2/2 q^2- \hbar \omega /2- \mu_\sigma)\right) \right] \\-
\left.
\log\left[1 + \exp\left(\beta(\epsilon_\sigma(q/2) + m_\sigma \omega^2/2 q^2 + \hbar \omega /2- \mu_\sigma)\right) \right]
\right\}.
\end{multline}
In the two-dimensional case, we must resort to numerical methods to determine $\Im m \Pi_\sigma$ at arbitrary temperatures. Substitution of the result for $\Im m \Pi_\sigma$ into Eq.~(\ref{eq:gadh}) leads to our final expression for $\Gamma^{\rm D}$ after decoupling in the Hartree channel. The real part of $\Pi_\sigma$, which is typically present in the interaction $W$, can be obtained by a Kramers-Kronig transform of $\Im m \Pi_\sigma$. We note that since $\Pi$ is an intra-species quantity, the presence of imbalance in the chemical potential or mass does not make the determination of $\Pi_\sigma$ more difficult as compared to the balanced case.

\subsection{Cooper instability}

When the effective interaction $W$ depends strongly on the center-of-mass momentum ${\bm K} = {\bm k}_1 + {\bm k}_2$ and total energy $\hbar \omega = \xi_1 + \xi_2$ of the incoming particles, the Cooper channel is the most appropriate channel to decouple the collision integral in Eq.~(\ref{eq:gadq}). In this case we thus introduce the energy variable $\hbar \omega$  using the identity
\ber
\delta(\xi_1 + \xi_2 - \xi_3 - \xi_4) &=& \int d(\hbar \omega)~\delta(\xi_1 + \xi_2 - \hbar \omega) \nonumber \\
&\times &\delta(\xi_3 + \xi_4 - \hbar \omega)~,
\eer
and solve the delta function of momentum conservation by setting ${\bm k}_1 = {\bm K}/2 + {\bm k}$, ${\bm k}_2 = {\bm K}/2 - {\bm k}$, ${\bm k}_3 = {\bm K}/2 - {\bm k}'$, and ${\bm k}_4 = {\bm K}/2 + {\bm k}'$. We find:
\begin{multline}  \label{eq:gadc}
\Gamma^{\rm D} = s_\uparrow s_\downarrow
\frac{\pi \beta \hbar^2}{d  V^3 m_\uparrow m_\downarrow} \sum_{{\bm k}, {\bm k}', {\bm K}}  |W({\bm K},\hbar \omega)|^2
({\bm k}'-{\bm k})^2 \\
[\delta(\xi_\uparrow({\bm K}/2 + {\bm k})+\xi_\downarrow({\bm K}/2 - {\bm k}) - \hbar \omega)  \\  n_{\rm F}(\xi_\uparrow({\bm K}/2 + {\bm k})) n_{\rm F}(\xi_\downarrow({\bm K}/2 - {\bm k})) ] \\
\{\delta(\xi_\uparrow({\bm K}/2 + {\bm k}')+\xi_\downarrow({\bm K}/2 - {\bm k}') - \hbar \omega) \\
[1-n_{\rm F}(\xi_\uparrow({\bm K}/2 + {\bm k}'))]
[1- n_{\rm F}(\xi_\downarrow({\bm K}/2 - {\bm k}'))]\}~.
\end{multline}
Expanding the factor $({\bm k}'-{\bm k})^2$ we can decouple the ${\bm k}$ and ${\bm k}'$ summations and obtain
\begin{multline} \label{eq:gadc2}
\Gamma^{\rm D} = s_\uparrow s_\downarrow
\frac{\beta \hbar^2}{2 \pi d V m_\uparrow m_\downarrow} \int d(\hbar \omega) \sum_{{\bm K}}  \frac{|W({\bm K},\hbar \omega)|^2}{\sinh^2(\beta \hbar \omega/2)}\\ [\Im m~\Xi_0({\bm K}, \hbar \omega) \Im m~\Xi_2({\bm K},\hbar \omega)  - \Im m~\Xi_1^2({\bm K},\hbar \omega)],
\end{multline}
where we have introduced the following three ``generalized pairing susceptibilities" ($n=0,1$, and $2$)
\begin{multline} \label{eq:imxn}
\Im m~\Xi_n({\bm K},\hbar\omega) =  \frac{1}{V} \Im m \sum_{{\bm k}} C_n({\bm k})  \\ \frac{1-n_{\rm F}(\xi_\uparrow({\bm K}/2+{\bm k}))- n_{\rm F}(\xi_\downarrow({\bm K}/2-{\bm k}))}{\hbar \omega + i 0^+  - (\xi_\uparrow({\bm K}/2+{\bm k}) + \xi_\downarrow({\bm K}/2-{\bm k}))}~.
\end{multline}
Here $C_0({\bm k}) = 1$, $C_1({\bm k})= {\bm k}$, and $C_2({\bm k}) = |{\bm k}|^2$. Note that we introduced a minor abuse of notation for the sake of accessibility of our formalism: although  $\Im m\Xi_1$ and $C_1$ are vector valued functions, they are not denoted by a bold-faced symbol. In determining the susceptibility $\Xi$, we restrict ourselves to the case of bands of equal character $s_\uparrow = s_\downarrow = 1$. It will turn out that the case $s_\uparrow = - s_\downarrow$ can be obtained by considering the susceptibility $\Delta$ defined in Eq.~(\ref{eq:de}), obtained by performing the Fock decoupling in the case that the bands have equal character $s_\uparrow = s_\downarrow$. We will show this point explicitly in the next section in which we perform the Fock decoupling.

To determine the generalized susceptibility it is convenient to make a shift of the summation variable ${\bm k}$ such that the denominator in Eq.~(\ref{eq:imxn}) becomes independent of the angle $\theta$ between ${\bm k}$ and ${\bm K}$.\footnote{The shifted ${\bm k}$ is the relative momentum, defined as the relative velocity times the effective mass $\frac{m_\uparrow m_\downarrow}{m_\uparrow+m_\downarrow}$ of the two species.  Then the kinetic energy of the pair is simply the sum of the energy of the center of mass and the relative energy.}  We do not need to shift the ${\bm k}$ as argument of the $C_n$'s, as can be seen by making the shift for both ${\bm k}$ and ${\bm k}'$ in Eq.~(\ref{eq:gadc}) before expanding the factor $({\bm k}'-{\bm k})^2$. The result is
\begin{multline}
\Im m\Xi_n({\bm K},\hbar\omega)  =  \frac{1}{V} \Im m \sum_{{\bm k}}  C_n({\bm k})  \\  \frac{1-n_{\rm F}(\xi_\uparrow(R_\uparrow {\bm K}+{\bm k}))- n_{\rm F}(\xi_\downarrow(R_\downarrow {\bm K}-{\bm k}))}{\hbar \omega + i 0^+  - (\xi_\uparrow(R_\uparrow {\bm K}+{\bm k}) + \xi_\downarrow(R_\downarrow {\bm K}-{\bm k}))},
\end{multline}
where $R_\sigma = m_\sigma / (m_\uparrow + m_\downarrow)$ and $R_\uparrow + R_\downarrow = 1$. We convert the summation to an integration and introduce the delta function
\begin{multline}
\Im m\Xi_n({\bm K},\hbar\omega)   =  - \frac{\pi}{(2 \pi)^d} \Im m \int d {\bm k}  C_n({\bm k})
\\ \left[1-n_{\rm F}(\xi_\uparrow(R_\uparrow {\bm K}+{\bm k}))- n_{\rm F}(\xi_\downarrow(R_\downarrow {\bm K}-{\bm k}))\right]\\
\delta(\mu_\uparrow + \mu_\downarrow + \hbar \omega   - (\hbar^2 k^2/2 m_\uparrow + \hbar^2 k^2/2 m_\downarrow +  \hbar^2 K^2/2 (m_\uparrow + m_\downarrow))),
\end{multline}
where we choose ${\bm K}$ along the $x$-axis in the two-dimensional case and ${\bm K}$ along the $z$-axis in the three-dimensional case, and where we noted that $\Im m\Xi_1({\bm K},\hbar\omega)$  can only have a component in the direction of ${\bm K}$. The solution of the delta function is
\be
k_0 = \frac{1}{\sqrt{\hbar^2 /2 m_\uparrow + \hbar^2/2 m_\downarrow}}\sqrt{\mu_\uparrow + \mu_\downarrow + \hbar \omega -  \frac{\hbar^2 K^2}{2 (m_\uparrow + m_\downarrow)}}~.
\ee
This solution only exists when
\be
K^2 \leq  \frac{\mu_\uparrow + \mu_\downarrow + \hbar \omega}{\hbar^2/2 (m_\uparrow + m_\downarrow)} \equiv K^2_\text{max},
\ee
which gives an upper bound for the length of ${\bm K}$ in Eq.~(\ref{eq:gadc2}). For $K> K_{\rm max}$ the imaginary part of $\Xi$ is zero. We solve the delta function and take into account the Jacobian and the transformation rule for the delta function, which leads for the three-dimensional case to
\begin{widetext}
\be
\Im m\Xi_n({\bm K},\hbar\omega) =
 \frac{-\pi}{(2 \pi)^2} \int_0^\pi  d \theta
 C_n({\bm k})
\frac{ k^2_0 \sin(\theta)}{2 k_0 (\hbar^2 /2 m_\uparrow + \hbar^2/2 m_\downarrow)}  [1 - n_{\rm F}(E_\uparrow(\theta)) - n_{\rm F}(E_\downarrow(\theta))].
\ee
The expression for the two-dimensional case differs in the Jacobian and the upper-integration boundary for $\theta$:
\be
\Im m\Xi_n({\bm K},\hbar\omega)  =
\frac{-\pi}{(2 \pi)^2} \int_0^{2\pi}  d \theta
 C_n({\bm k})
\frac{ k_0}{2 k_0 (\hbar^2 /2 m_\uparrow + \hbar^2/2 m_\downarrow)}   [1 - n_{\rm F}(E_\uparrow(\theta)) - n_{\rm F}(E_\downarrow(\theta))].
\ee
In the previous equation, we defined the energies
\be
E_\uparrow(\theta) = \hbar^2 k_0^2/2 m_\uparrow + m_\uparrow \hbar^2 K^2/2 (m_\uparrow + m_\downarrow)^2 + \cos(\theta) \hbar^2 k_0 K /2 (m_\uparrow + m_\downarrow) - \mu_\uparrow
\ee
and
\be
E_\downarrow(\theta) = \hbar^2 k_0^2/2 m_\downarrow + m_\downarrow \hbar^2 K^2/2 (m_\uparrow + m_\downarrow)^2 - \cos(\theta) \hbar^2 k_0 K /2 (m_\uparrow + m_\downarrow) - \mu_\downarrow.
\ee
\end{widetext}
Using the integrals
\be
\int_0^\pi d \theta \frac{\sin(\theta)}{1 + \exp(a + b\cos(\theta))} = 2 + \frac{1}{b} \log\left[\frac{1+\exp(a-b)}{1+\exp(a+b)} \right],
\ee
and
\begin{multline} \label{eq:polylocation}
\int_0^\pi d \theta \frac{\cos(\theta) \sin(\theta)}{1 + \exp(a + b\cos(\theta))} \\=
- \frac{1}{b} \log\left[(1+\exp(a-b))(1+\exp(a+b))\right]+ \\
\frac{1}{b^2}\left(\text{Li}_2(-\exp(a-b)) -\text{Li}_2(-\exp(a+b))\right),
\end{multline}
where $\text{Li}_s(z) = \sum_{k=1}^\infty z^k/k^s$ is the polylogarithm, we can determine the susceptibility $\Im m\Xi_n({\bm K},\hbar\omega)$ for the three-dimensional case in closed form. However, the resulting expressions are not very enlightening and will not be given explicitly. Substitution of the results for $\Im m \Xi_n$ into Eq.~(\ref{eq:gadc2}) leads to our final expression for $\Gamma^{\rm D}$ after decoupling in the Cooper channel.

\subsection{Fock instability}
\label{sec:qf}
When the interaction depends strongly on the energy and momentum difference of the incoming $\uparrow$ particle and outgoing $\downarrow$ particle, the Fock channel is the appropriate channel to decouple the collision integral Eq.~(\ref{eq:gadq}). We introduce a new energy variable $\hbar \omega = \xi_1 - \xi_3$  using the identity
\be
\delta(\xi_1 + \xi_2 - \xi_3 - \xi_4) = \int d \hbar \omega \delta(\xi_1 - \xi_3 - \hbar \omega) \delta(\xi_2 - \xi_4 + \hbar \omega),
\ee
and the ``conjugate" momentum variable ${\bm Q}= {\bm k}_1 - {\bm k}_3$ by solving the momentum-conserving delta function by ${\bm k}_1 = {\bm k}+{\bm Q}$, ${\bm k}_2 = {\bm k}'$, ${\bm k}_3 = {\bm k}$, and ${\bm k}_4= {\bm k}'+{\bm Q}$. Then, we have for $\Gamma^{\rm D}$
\begin{multline}
\Gamma^{\rm D} = s_\uparrow s_\downarrow  \frac{\pi \beta \hbar^2}{d V^3 m_\uparrow m_\downarrow}   \sum_{{\bm k},{\bm k}',{\bm Q}} \int d \hbar \omega |W({\bm Q},\hbar \omega)|^2   ({\bm k}' -{\bm k})^2 \\
\delta(\xi_\uparrow({\bm k}+{\bm Q}) - \xi_\downarrow({\bm k}) - \hbar \omega) n_{\rm F}(\xi_\uparrow({\bm k}+{\bm Q})) (1-n_{\rm F}(\xi_\downarrow({\bm k})))  \\
\delta(\xi_\downarrow({\bm k}') - \xi_\uparrow({\bm k}'+{\bm Q}) + \hbar \omega) n_{\rm F}(\xi_\downarrow({\bm k}')) (1-n_{\rm F}(\xi_\uparrow({\bm k}'+{\bm Q}))).
\end{multline}
We again expand the factor $({\bm k}' -{\bm k})^2$ and obtain
\begin{multline} \label{eq:gadfoq}
\Gamma^{\rm D} = s_\uparrow s_\downarrow  \frac{\beta \hbar^2}{2 \pi d  V m_\uparrow m_\downarrow}   \sum_{{\bm Q}} \int d \hbar \omega
\frac{|W({\bm Q},\hbar \omega)|^2 }{\sinh^2(\beta \hbar \omega/2)} \\ [ \Im m\Delta_0({\bm Q},\hbar \omega) \Im m\Delta_2({\bm Q},\hbar \omega) - \Im m\Delta^2_1({\bm Q},\hbar \omega)],
\end{multline}
where we defined:
\begin{multline} \label{eq:de}
\Im m\Delta_n({\bm Q},\hbar\omega) = \frac{1}{V}  \Im m \sum_{{\bm k}}  F_n({\bm k})  \\ \frac{n_{\rm F}(\xi_\uparrow({\bm k}+{\bm Q}))- n_{\rm F}(\xi_\downarrow({\bm k}))}{\xi_\uparrow({\bm k}+{\bm Q}) - \xi_\downarrow({\bm k})- \hbar \omega - i 0^+}.
\end{multline}
Here $F_0({\bm k}) = 1$, $F_1({\bm k})= {\bm k}$, and $F_2({\bm k}) = |{\bm k}|^2$. Now, we compare the susceptibilities for the Cooper and the Fock channel, $\Xi$ and $\Delta$, respectively. Defining $\tilde{\xi}_\downarrow(k) = -\xi_\downarrow(k)$ we rewrite the fraction in Eq.~(\ref{eq:de}) as
\be
\frac{1-n_{\rm F}(\xi_\uparrow({\bm k}+{\bm Q}))- n_{\rm F}(\tilde{\xi}_\downarrow({\bm k}))}{\hbar \omega + i 0^+-\xi_\uparrow({\bm k}+{\bm Q}) - \tilde{\xi}_\downarrow({\bm k})},
\ee
which is exactly the expression for $\Xi_n({\bm Q},\hbar\omega)$ in Eq.~(\ref{eq:imxn}) with a switched character for the $\downarrow$-dispersion. Thus, when determining $\Xi_n$ and $\Delta_n$, we can restrict ourselves to the case $s_\uparrow = s_\downarrow =1$. When we need $\Xi_n$ for dispersions of opposite character ($s_\uparrow=-s_\downarrow$), this is just obtained by considering $\Delta_n$ for bands of the same character ($s_\uparrow = s_\downarrow =1$).

Following up on the remarks made above, we now determine $\Delta_n ({\bf Q}, \hbar \omega)$ for bands of the same character $s_\uparrow = s_\downarrow =1$. The way the analysis proceeds depends on whether or not there is mass imbalance. The reason is that when $m_\uparrow = m_\downarrow$ the terms with $k^2$ cancel from the denominator of the integrand of Eq.~(\ref{eq:de}), whereas they do not cancel in the mass-imbalanced case. For sake of simplicity, we  only consider the fully balanced case with  $m_\uparrow = m_\downarrow \equiv m$ and $\mu_\uparrow = \mu_\downarrow \equiv \mu$. Shifting ${\bm k}$ with ${\bm Q}/2$, we arrive at
\begin{multline}
\Im m\Delta_n({\bm Q},\hbar\omega) = \frac{\pi}{(2 \pi)^d}  \Im m \int d {\bm k}  F_n({\bm k}) \\ [n_{\rm F}(\xi_\uparrow({\bm k}+{\bm Q}/2))- n_{\rm F}(\xi_\downarrow({\bm k}-{\bm Q}/2))] \delta((\hbar^2 k Q/m) \cos(\theta) - \hbar \omega),
\end{multline}
where $\theta$ is the angle between ${\bm k}$ and ${\bm Q}$. In three dimensions, we choose ${\bm Q}$ along the $z$-axis, so that $\theta$ is the polar angle of ${\bm k}$. In two dimensions, we choose ${\bm Q}$ along the $x$-axis so that $\theta$ is the angle between ${\bm k}$ and the $x$-axis. After some calculations, we arrive at the result
\begin{multline}
\Im m \Delta_n({\bm Q},\hbar \omega) = \frac{\pi}{(2 \pi)^d} \int_{k_{\text{min}}}^\infty d k F_{n,d}(k)
\\
\left[
n_{\rm F}\left(\epsilon(k) + \epsilon(Q/2) - \mu + \hbar \omega/2 \right)\right. \\ - \left.
n_{\rm F}\left(\epsilon(k) + \epsilon(Q/2) - \mu - \hbar \omega/2 \right)
\right],
\end{multline}
where $\epsilon(k) = \hbar^2 k^2/2m$ and where $k_{\text{min}} = m |\hbar \omega|/\hbar^2 Q$. The functions $F_{0,d}$ depend on dimensionality:
\be
F_{0,2}(k) = \frac{2 k}{\sqrt{\hbar^4 k^2 Q^2/ m^2 -\hbar^2 \omega^2}} \quad \text{and} \quad
F_{0,3}(k) = \frac{2 \pi k m}{\hbar^2 Q}.
\ee
Finally, $F_{1,d}(k)=(\omega m/\hbar Q) F_{0,d}(k) $, and $F_{2,d}(k) = k^2 F_{0,d}(k)$. In the three-dimensional case, we can find closed form expressions for $\Delta_n$ using the following integrals
\be
\int d k \frac{k}{1+\exp(a(b+k^2))} = \frac{k^2}{2} - \frac{1}{2a} \log(1+\exp(a(b+k^2))),
\ee
and
\begin{multline}
\int d k \frac{k^3}{1+\exp(a(b+k^2))} = \frac{k^4}{4} - \frac{k^2}{2a} \log(1+\exp(a(b+k^2))) \\ - \frac{1}{2a^2} \text{Li}_2(-\exp(a(b+k^2))).
\end{multline}
where $\text{Li}_s(z)$ the polylogarithm defined after Eq.~(\ref{eq:polylocation}). These expressions are, however, not particularly enlightening and will not be given explicitly here.

\section{Linear Dispersion}
\label{sec:lin}
For a linear dispersion, Eq.~(\ref{eq:gad}) simplifies to
\begin{multline} \label{eq:gadl}
\Gamma^{\rm D} = - s_\uparrow s_\downarrow \frac{1}{4} \frac{\pi \beta v^2}{d V^3} \sum_{{\bm k}_i}  \delta_{{\bm k}_i} \delta(\xi_i)
|W|^2 n_1 n_2 (1-n_3) (1- n_4) \\ (\hat{{\bm k}}_4 -\hat{{\bm k}}_1)  \cdot (\hat{{\bm k}}_3 - \hat{{\bm k}}_2)
\left(1+\hat{{\bm k}}_1 \cdot \hat{{\bm k}}_4  \right)
\left(1+\hat{{\bm k}}_2 \cdot \hat{{\bm k}}_3  \right)~,
\end{multline}
where we reinstated the chirality factors. The main purpose of this section is to derive the expressions that were the starting point of Ref.~[\onlinecite{mink}], where the drag resistivity was determined in a topological insulator thin film close to exciton condensation. Then, $s_\uparrow =1$ and $s_\uparrow =-1$ and the appropriate decoupling channel for this transition is the Fock channel. For comparison, we also perform the decoupling of Eq.~(\ref{eq:gadl}) in the Hartree channel.

In Ref.~[\onlinecite{mink}], we used an approximation for $\Gamma^{\rm D}$ in which the chirality factors $\left(1+\hat{{\bm k}}_1 \cdot \hat{{\bm k}}_4  \right)\left(1+\hat{{\bm k}}_2 \cdot \hat{{\bm k}}_3  \right)$  were set to unity. The effect of the chirality factors is the suppression of backscattering: $1+\hat{{\bm k}}_1 \cdot \hat{{\bm k}}_4 $ vanishes when ${\bm k}_1$ and ${\bm k}_4$ are directed oppositely. The pole in the scattering amplitude $W$ for the case of exciton condensation occurs at ${\bm Q} = {\bm k}_1 - {\bm k}_3 =0 $. The motivation for this approximation is that the condition for backscattering $\hat{{\bm k}}_1 \cdot \hat{{\bm k}}_4=-1$ is largely independent of the value of ${\bm Q}$. In other words, the suppression of backscattering does not favor or suppress a particular value for the exciton momentum ${\bm Q}$. In particular, discarding the chirality factors will not have a qualitative influence on the behavior of $\Gamma^{\rm D}$ close to exciton condensation, which is determined by the ${\bm Q}$ dependence of the integrand of Eq.~(\ref{eq:gadl}).

\subsection{Hartree Instability}
When the interaction depends only on the transferred energy $\hbar \omega = \xi_1 - \xi_4$ and momentum ${\bm k}_p = {\bm k}_1 - {\bm k}_4 \equiv {\bm q}$ in the scattering event, the Hartree channel is the appropriate instability channel. We introduce an additional integral over $\hbar \omega$ using the identity
\be
\delta(\xi_1 + \xi_2 - \xi_3 - \xi_4) = \int d \hbar \omega \delta(\xi_1 - \xi_4 - \hbar \omega) \delta(\xi_2 - \xi_3 + \hbar \omega),
\ee
and solve the momentum-conserving delta function by ${\bm k}_1 = {\bm k}$, ${\bm k}_2 = {\bm k}'$, ${\bm k}_3 = {\bm k}'+{\bm q}$, and ${\bm k}_4={\bm k}-{\bm q}$, so that the ${\bm k}$ and ${\bm k}'$ summations in Eq.~(\ref{eq:gadl}) decouple
\begin{multline}
\Gamma^{\rm D} = - s_\uparrow s_\downarrow \frac{\pi \beta  v^2}{d  V^3} \sum_{{\bm k},{\bm k}',{\bm q}}   \int d \hbar \omega |W({\bm q}, \hbar \omega)|^2 \\
\delta(\xi_\uparrow({\bm k}) - \xi_\uparrow({\bm k}-{\bm q}) - \hbar \omega) n_{\rm F}(\xi_\uparrow({\bm k})) (1- n_{\rm F}(\xi_\uparrow({\bm k}-{\bm q})))
\\ \left(\frac{{\bm k} - {\bm q}}{|{\bm k} - {\bm q}|} -\hat{{\bm k}}\right) \cdot  \left(\frac{{\bm k}' + {\bm q}}{|{\bm k}' + {\bm q}|} -\hat{{\bm k}'}\right) \\ \delta(\xi_\downarrow({\bm k}') - \xi_\downarrow({\bm k}'+{\bm q}) + \hbar \omega) n_{\rm F}(\xi_\downarrow({\bm k}')) (1-n_{\rm F}(\xi_\downarrow({\bm k}'+{\bm q}))).
\end{multline}
This expression can be rewritten as
\begin{multline}
\Gamma^{\rm D} =  s_\uparrow s_\downarrow \frac{\beta  v^2}{4 \pi  d  V} \sum_{{\bm k},{\bm k}',{\bm q}} \frac{ |W({\bm q}, \hbar \omega)|^2}{\sinh^2(\beta\hbar \omega/2)} \\
\Im m \Pi_\uparrow({\bm q},\hbar \omega) \cdot \Im m \Pi_\downarrow({\bm q},\hbar \omega),
\end{multline}
where we defined the polarizability
\begin{multline}
\Pi_\sigma({\bm q},\hbar \omega) = \frac{1}{V} \sum_{{\bm k}}  \frac{n_{\rm F}(\xi_\sigma({\bm k}+{\bm q})) - n_{\rm F}(\xi_\sigma({\bm k}))}{\xi_\sigma({\bm k}+{\bm q}) - \xi_\sigma({\bm k}) - \hbar \omega-i0}  \\  \left(\frac{{\bm k} + {\bm q}}{|{\bm k} + {\bm q}|} -\hat{{\bm k}}\right),
\end{multline}
and used that $\Pi_\sigma({\bm q},\omega) = -\Pi_\sigma(-{\bm q},\omega)$ and $\Pi_\sigma({\bm q},-\omega) = -\Pi^*_\sigma({\bm q},\omega)$. Using that $\Pi_\sigma({\bm q},\omega)$ is independent of $s_\sigma$ we find
\begin{multline}
\Pi_\sigma({\bm q},\hbar \omega) = \frac{1}{2 \pi} \int_0^\pi d \theta  \int_0^\infty k dk  \\ \left(\frac{q + k \cos(\theta)}{\sqrt{q^2 + k^2 + 2 k q \cos(\theta)}} - \cos(\theta)\right) \\  (n_{\rm F}(\hbar v \sqrt{k^2 + q^2 + 2 k q \cos(\theta)}-\mu_\sigma) - n_{\rm F}(\hbar v k - \mu_\sigma)) \\ \delta(\hbar v \sqrt{k^2 + q^2 + 2 k q \cos(\theta)} - \hbar v k - \hbar \omega),
\end{multline}
where we restricted the angular integration between $0$ and $\pi$. The delta function has the solution $k = k_0$ with
\be
k_0 = \frac{\omega^2-v^2 q^2}{2 v (q v \cos(\theta)-\omega)}
\ee
when $|\omega|/vq<1$ and $\cos^{-1}(\omega/v)<\theta\equiv \theta_{\rm min}$. After solving the delta function, $\Pi_\sigma$ becomes
\begin{multline}
\Pi_\sigma({\bm q},\hbar \omega) = \frac{\hbar}{2 \pi} \frac{(v q)^2 - (\omega)^2}{2  \epsilon_{\rm F}} \int_{\theta_{min}}^\pi d \theta \frac{q v - \omega \cos(\theta)}{(\omega - q v \cos(\theta))^2 } \\
\left\{
n_{\rm F}\left[\left(\frac{\hbar(\omega^2-q^2 v^2)}{2 (q v \cos(\theta)-\omega)}-\mu_\sigma+\hbar\omega\right)/T\right] \right.\\
\left.-n_{\rm F}\left[\left(\frac{\hbar(\omega^2-q^2 v^2)}{2 (q v \cos(\theta)-\omega)}-\mu_\sigma\right)/T\right] \right\}.
\end{multline}
Substitution of the result for $\Im m \Pi_\sigma$ into Eq.~(\ref{eq:gadl}) leads to our final expression for $\Gamma^{\rm D}$ after decoupling in the Hartree channel.

\subsection{Fock Instability}
When the interaction depends mainly on the energy and momentum difference of the incoming $\uparrow$ particle and outgoing $\downarrow$ particle, the Fock channel is the appropriate channel to decouple the collision integral Eq.~(\ref{eq:gadq}). We introduce the energy variable $\hbar \omega = \xi_1 - \xi_3$ using the identity
\be
\delta(\xi_1 + \xi_2 - \xi_3 - \xi_4) = \int d \hbar \omega \delta(\xi_1 - \xi_3 - \hbar \omega) \delta(\xi_2 - \xi_4 + \hbar \omega),
\ee
and the momentum variable ${\bm Q}= {\bm k}_1 - {\bm k}_3$ by solving the momentum-conserving delta function by ${\bm k}_1 = {\bm k}+{\bm Q}$, ${\bm k}_2 = {\bm k}'$, ${\bm k}_3 = {\bm k}$, and ${\bm k}_4= {\bm k}'+{\bm Q}$. Then, we have for $\Gamma^{\rm D}$
\begin{multline} \label{eq:gadfoli}
\Gamma^{\rm D} = s_\uparrow s_\downarrow \frac{\pi \beta  v^2}{ d  V^3}  \sum_{{\bm k},{\bm k}',{\bm Q}} \int d \hbar \omega |W({\bm Q},\hbar \omega)|^2\\
\left(\frac{{\bm k}'+{\bm Q}}{|{\bm k}'+{\bm Q}|}-\frac{{\bm k}+{\bm Q}}{|{\bm k}+{\bm Q}|}\right)  \cdot (\hat{{\bm k}}' - \hat{{\bm k}}) \\
\delta(\xi_\uparrow({\bm k}+{\bm Q}) - \xi_\downarrow({\bm k}) - \hbar \omega) n_{\rm F}(\xi_\uparrow({\bm k}+{\bm Q})) (1-n_{\rm F}(\xi_\downarrow({\bm k})))  \\
\delta(\xi_\downarrow({\bm k}') - \xi_\uparrow({\bm k}'+{\bm Q}) + \hbar \omega) n_{\rm F}(\xi_\downarrow({\bm k}')) (1-n_{\rm F}(\xi_\uparrow({\bm k}'+{\bm Q}))).
\end{multline}
Expanding the inner product between the unit vectors, we obtain after some rewriting
\begin{multline} \label{eq:gadfol}
\Gamma^{\rm D} = s_\uparrow s_\downarrow \frac{\beta  v^2}{2 \pi d  V}  \sum_{{\bm k},{\bm k}',{\bm Q}} \int d \hbar \omega |W({\bm Q},\hbar \omega)|^2 \\
[\Im m\Delta_0({\bm Q},\hbar \omega) \Im m\Delta_2({\bm Q},\hbar \omega)  \\ - \Im m\Delta_{1a}({\bm Q},\hbar \omega) \cdot \Im m \Delta'_{1b}({\bm Q},\hbar \omega)].
\end{multline}
where we have defined the generalized susceptibility
\begin{multline} \label{eq:den1}
\Im m\Delta_n({\bm Q},\hbar \omega)  = \frac{1}{V} \Im m \sum_{{\bm k}} F_n({\bm k})\\
\frac{n_{\rm F}(\xi_\uparrow({\bm k}+{\bm Q})) -n_{\rm F}(\xi_\downarrow({\bm k}))}{\xi_\uparrow({\bm k}+{\bm Q})-\xi_\downarrow({\bm k})  - \hbar \omega - i 0},
\end{multline}
where $F_0({\bm k}) = 1$, $F_{1a}({\bm k}) = \hat{{\bm k}}$, $F_{1b}({\bm k}) = ({\bm k}+{\bm Q})|{\bm k}+{\bm Q}|$, $F_{2}({\bm k}) = \hat{{\bm k}} \cdot ({\bm k}+{\bm Q})|{\bm k}+{\bm Q}|$. The reason that these $F$'s are more complicated than for the quadratic case is that the exciton momentum ${\bm Q}$ does not cancel from the inner product in Eq.~(\ref{eq:gadfoli}). We note that from Eq.~(\ref{eq:gadfol}) we obtain Eq.~(3) of Ref.~[\onlinecite{mink}]. We evaluate $\Im m\Delta_n$ for $s_\uparrow = 1$ and $s_\downarrow = - 1$, as is appropriate for the case of exciton condensation. Evaluating the imaginary part of Eq.~(\ref{eq:den1}), we find
\begin{multline} \label{eq:den2}
\Im m\Delta_n({\bm Q},\hbar \omega)  = -\frac{1}{4 \pi} \int k d k  \int_0^{2\pi} d \theta F_n(k,\theta)\\
(1 - n_{\rm F}(\hbar v \sqrt{k^2 + Q^2 + 2 k Q \cos(\theta)} - \mu_\uparrow) - n_{\rm F}(\hbar v k - \mu_\downarrow))\\
\delta(\hbar \omega + \mu_\uparrow + \mu_\downarrow - \hbar v k - \hbar v \sqrt{k^2 + Q^2 + 2 k Q \cos(\theta)}).
\end{multline}
The solution of the delta function is
\be
k_0 = \frac{(\hbar \omega + \mu_\uparrow + \mu_\downarrow)^2 - (\hbar v Q)^2}{2 \hbar v (\hbar \omega + \mu_\uparrow + \mu_\downarrow + \hbar v Q \cos(\theta))},
\ee
when $\hbar \omega + \mu_\uparrow + \mu_\downarrow > \hbar v Q$. Solving the delta function in Eq.~(\ref{eq:den2}) leads to
\begin{multline} \label{eq:den3}
\Im m\Delta_n({\bm Q},\hbar \omega)  =  \frac{(\hbar \omega + \mu_\uparrow + \mu_\downarrow)^2 - (\hbar v Q)^2}{8 \pi (\hbar v)^2} \int_0^{\pi} d \theta F_n(\theta)\\
\frac{(\hbar \omega + \mu_\uparrow + \mu_\downarrow)^2 + (\hbar v Q)^2+2 \hbar v Q (\hbar \omega + \mu_\uparrow + \mu_\downarrow)\cos(\theta)}{[(\hbar \omega + \mu_\uparrow + \mu_\downarrow) + \hbar v Q \cos(\theta)]^3}\\
\left[1 - n_{\rm F}\left(-\frac{(\hbar \omega + \mu_\uparrow + \mu_\downarrow)^2 - (\hbar v Q)^2}{2[(\hbar \omega + \mu_\uparrow + \mu_\downarrow) + \hbar v Q \cos(\theta)]} + \mu_\downarrow + \hbar \omega \right) - \right. \\ \left.
n_{\rm F}\left(\frac{(\hbar \omega + \mu_\uparrow + \mu_\downarrow)^2 - (\hbar v Q)^2}{2[(\hbar \omega + \mu_\uparrow + \mu_\downarrow) + \hbar v Q \cos(\theta)]} - \mu_\downarrow\right) \right].
\end{multline}
The functions $F_n(\theta)$ are given by $F_0(\theta) = 1$, $F_{1a}(\theta) = \cos(\theta)$,
\begin{multline}
F_{1b}(\theta) = \\\frac{[(\hbar \omega + \mu_\uparrow + \mu_\downarrow)^2 + (\hbar v Q)^2]\cos(\theta)+2 \hbar v Q (\hbar \omega + \mu_\uparrow + \mu_\downarrow)}{(\hbar \omega + \mu_\uparrow + \mu_\downarrow)^2 + (\hbar v Q)^2+2 \hbar v Q (\hbar \omega + \mu_\uparrow + \mu_\downarrow)\cos(\theta)},
\end{multline}
and
\begin{widetext}
\begin{equation}
F_{2}(\theta) = \frac{(\hbar \omega + \mu_\uparrow + \mu_\downarrow)^2 + 2 \hbar v Q (\hbar \omega + \mu_\uparrow + \mu_\downarrow)\cos(\theta)+ (\hbar v Q)^2 \cos(2 \theta)}{(\hbar \omega + \mu_\uparrow + \mu_\downarrow)^2 + (\hbar v Q)^2+2 \hbar v Q (\hbar \omega + \mu_\uparrow + \mu_\downarrow)\cos(\theta)}.
\end{equation}
Substitution of the result for $\Im m \Pi_\sigma$ into Eq.~(\ref{eq:gadl}) leads to our final expression for $\Gamma^{\rm D}$ after decoupling in the Fock channel.\cite{mink}
\end{widetext}

\section{Basic applications}
\label{sec:allresults}
Before we quantitatively (and therefore numerically) consider two basic applications of our theory, we present some qualitative results on the dependence of the drag resistivity close to a phase transition.
These results depend only on (i) the assumed spectrum of Gaussian fluctuations close to the transition and (ii) power counting.   Therefore they are very general.
\subsection{Qualitative results}
\label{subsec:qualitative}
Consider first the Hartree channel in three dimensions. At low temperatures we have in the balanced case that $\Im m \Pi_\sigma({\bm q},\hbar \omega) \propto \omega/q$, so that the leading order low-temperature behavior of the drag resistivity is determined by the integral
\begin{equation}
\label{eq:lowtrho}
 \rho_{\rm D} \propto \int d \omega \int d q q^4 \left[\frac{\omega }{q{\rm sinh} \left(\frac{ \beta \hbar \omega}{2}\right)}\right]^2 w (q,\omega)~,
\end{equation}
where the relevant momentum and energy dependence that results from the scattering amplitude (derived in detail below), is given by
\begin{equation}
 w(q,\omega) \propto \frac{1}{\alpha(T)+\left(\frac{q}{k_{\rm F}}\right)^2+\left(\frac{c_1 \omega}{qv_{\rm F}}\right)^2}~,
\end{equation}
where $\hbar k_{\rm F}$ and $v_{\rm F}$ are, respectively,  the Fermi momentum and the Fermi velocity, and $c_1$ is a constant. The important temperature dependence is determined by $\alpha (T) \propto T-T_{\rm c}$, which approaches zero as $T$ approaches the critical temperature $T_{\rm c}$ from above. At low temperatures only small energies are relevant: we therefore expand ${\rm sinh} \left( \beta \hbar \omega/2 \right) \simeq \beta \hbar \omega/2$. The frequency integral is then carried out and we find that the low-temperature behavior of the drag resistivity is determined by the integral
\begin{equation}
\label{eq:lowtrhoonlymomentum}
 \rho_{\rm D} \propto \int d q \frac{q^k}{\alpha (T) + c_2 q^2},
\end{equation}
with $c_2>0$ a constant independent of temperature, and $k=3$ for the Hartree channel in three dimensions. Carrying out the remaining momentum integral for $k=3$ leads to the conclusion that the drag resistivity remains finite within the Boltzmann theory presented in this paper.

For the Fock channel in three dimensions, a similar calculation leads to an expression of the form in Eq.~(\ref{eq:lowtrhoonlymomentum}) with $k=1$, so that the drag resistivity in that case diverges logarithmically. The difference between Hartree and Fock channels is understood from the fact that the  momentum ${\bf q}$ that controls  the critical fluctuations in the Hartree channel is precisely the momentum transferred in the collision, which leads to an additional factor $q^2$ in the integral in Eq.~(\ref{eq:lowtrhoonlymomentum}).  This is absent in the Fock channel, where the critical momentum {\bf Q} is different from the momentum transferred in the collisions, ${\bf k'}-{\bf k}$ (see Fig.~\ref{fig:Channels}).

For the Cooper channel in three dimensions the critical momentum ${\bf K}$ is again different from the momentum transfer $\kv'-\kv$ (see Fig. 2).  However, in this case the generalized susceptibilities are independent of momentum in the low-frequency limit: $\Im m \Xi({\bm q},\hbar \omega) \propto \omega$ -- a fact that compensates the ``loss" of the $q^2$ factor.   Furthermore, the critical scattering amplitude is given by
$$
 w(q,\omega) \propto \frac{1}{\alpha(T)+\left(\frac{q}{k_{\rm F}}\right)^2+\left(\frac{c_3 \omega}{\epsilon_{\rm F}}\right)^2}~,
$$
where $\epsilon_{\rm F}$ is the Fermi energy and $c_3$ is a constant.  The presence of $\omega/\epsilon_{\rm F}$ rather than $\omega/qv_{\rm F}$ in the denominator of this expression causes one less power of $q$ to appear after the integral over frequency is done.  The final result is of the form of Eq.~(\ref{eq:lowtrhoonlymomentum}) with $k=2$ so that the drag resistivity, in that case, remains finite at $T_{\rm c}$.

The above discussion of the Cooper-channel instability was for three dimensions. In two dimensions,  we have one power of $q$ less so that $k=1$ and we again find a logarithmic divergence (irrespective of band dispersion) for the drag resistivity if the instability occurs in the Cooper channel. In the next two sections we find that our numerical results agree with these power-counting arguments.
\begin{figure}
\begin{center}
\includegraphics[width = 0.5 \textwidth]{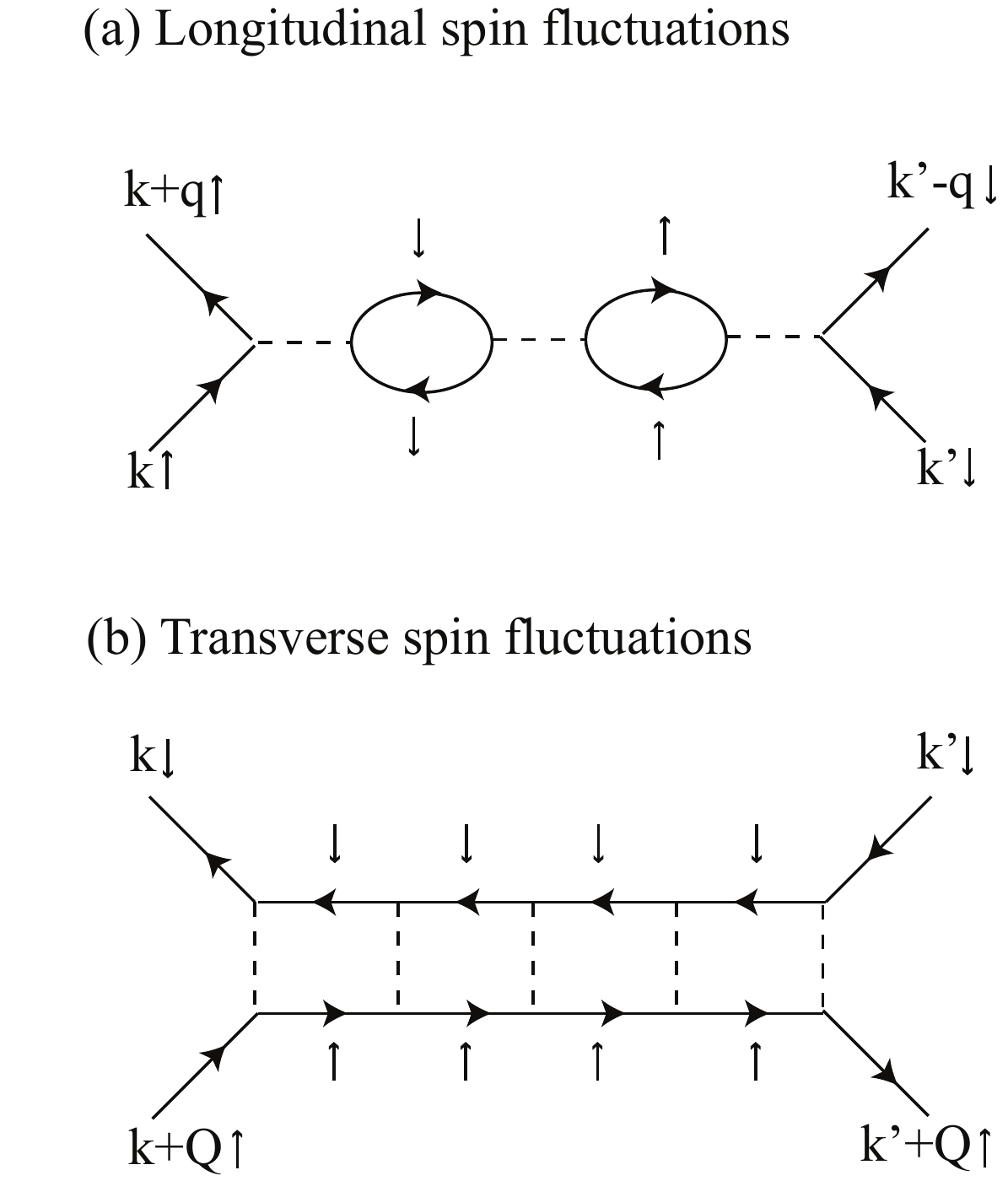}
\caption{ \label{fig:bubbles} Bubble Feynman diagrams that determine the dominant enhancement of the scattering amplitude near the ferromagnetic phase transition.}
\end{center}
\end{figure}
\subsection{Ferromagnetism in an ultracold Fermi gas}
\label{sec:res}
In this section we extend the results of Ref.~[\onlinecite{duinemag}], which considers the behavior of the spin-drag resistivity in an ultracold Fermi gas with repulsive interactions close to the ferromagnetic transition. This transition is accompanied by magnetic fluctuations of two different characters, longitudinal and transverse fluctuations. By summing all bubble diagrams that contribute to the interaction (see Fig.~\ref{fig:bubbles}) we obtain
\begin{multline}
|W({\bm k}_1,{\bm k}_2,{\bm k}_3,{\bm k}_4) |^2  \\ = |W_{\rm L}({\bm k}_1 - {\bm k}_4,\xi_1 - \xi_4)|^2 + |W_{\rm T}({\bm k}_1 - {\bm k}_3,\xi_1 - \xi_3)|^2,
\end{multline}
where $W_{\rm L}$ is the contribution from longitudinal fluctuations and $W_{\rm T}$ is the contribution from transverse fluctuations, which will be specified below. To determine the spin-drag resistivity we need to evaluate the collision integral in Eq.~(\ref{eq:gad}) for this scattering amplitude. We do this by splitting the collision integral into two integrals, one with scattering amplitude $W_{\rm L}$ and one with $W_{\rm T}$. The former integral can be evaluated by decoupling it in the Hartree channel as described in section~\ref{sec:qh}, and the latter by decoupling it in the Fock channel as described in section~\ref{sec:qf}. The final result for the drag resistivity is then the sum of these two contributions. In Ref.~[\onlinecite{duinemag}], the transverse fluctuations were included as if they were part of the Hartree channel. Instead, using the formalism described in this article, we can easily take into account the effect of both types of fluctuations.

The specific form of the scattering amplitude incorporating the longitudinal fluctuations is
\begin{multline}
W_{\rm L}({\bm q},\hbar \omega) =  U +
\frac{U^2}{4} \frac{\Pi({\bm q},\hbar \omega)}{1-U \Pi({\bm q},\hbar \omega)/2} \\
- \frac{U^2}{4} \frac{\Pi({\bm q},\hbar \omega)}{1+U \Pi({\bm q},\hbar \omega)/2},
\end{multline}
where $\Pi({\bm q},\hbar \omega) = \Pi_\uparrow({\bm q},\hbar \omega) + \Pi_\downarrow({\bm q},\hbar \omega)$ with $\Pi_\sigma({\bm q},\hbar \omega)$ given in Eq.~(\ref{eq:PI}), and where $U$ is the contact interaction strength given by $U = 4 \pi a \hbar^2/m$ with $a$ the $s$-wave scattering length. The specific form of the scattering amplitude incorporating the transverse fluctuations is
\begin{equation}
W_{\rm T}({\bm Q},\hbar \omega) = - \frac{U^2}{2} \frac{\Delta({\bm Q},\hbar \omega)}{1+U \Delta({\bm Q},\hbar \omega)/2},
\end{equation}
where $\Delta({\bm Q},\hbar \omega) = 2\Delta_0({\bm Q},\hbar \omega)$ with $\Delta_0$ given in Eq.~(\ref{eq:de}).

\begin{figure}
\begin{center}
\includegraphics[width = 0.5 \textwidth]{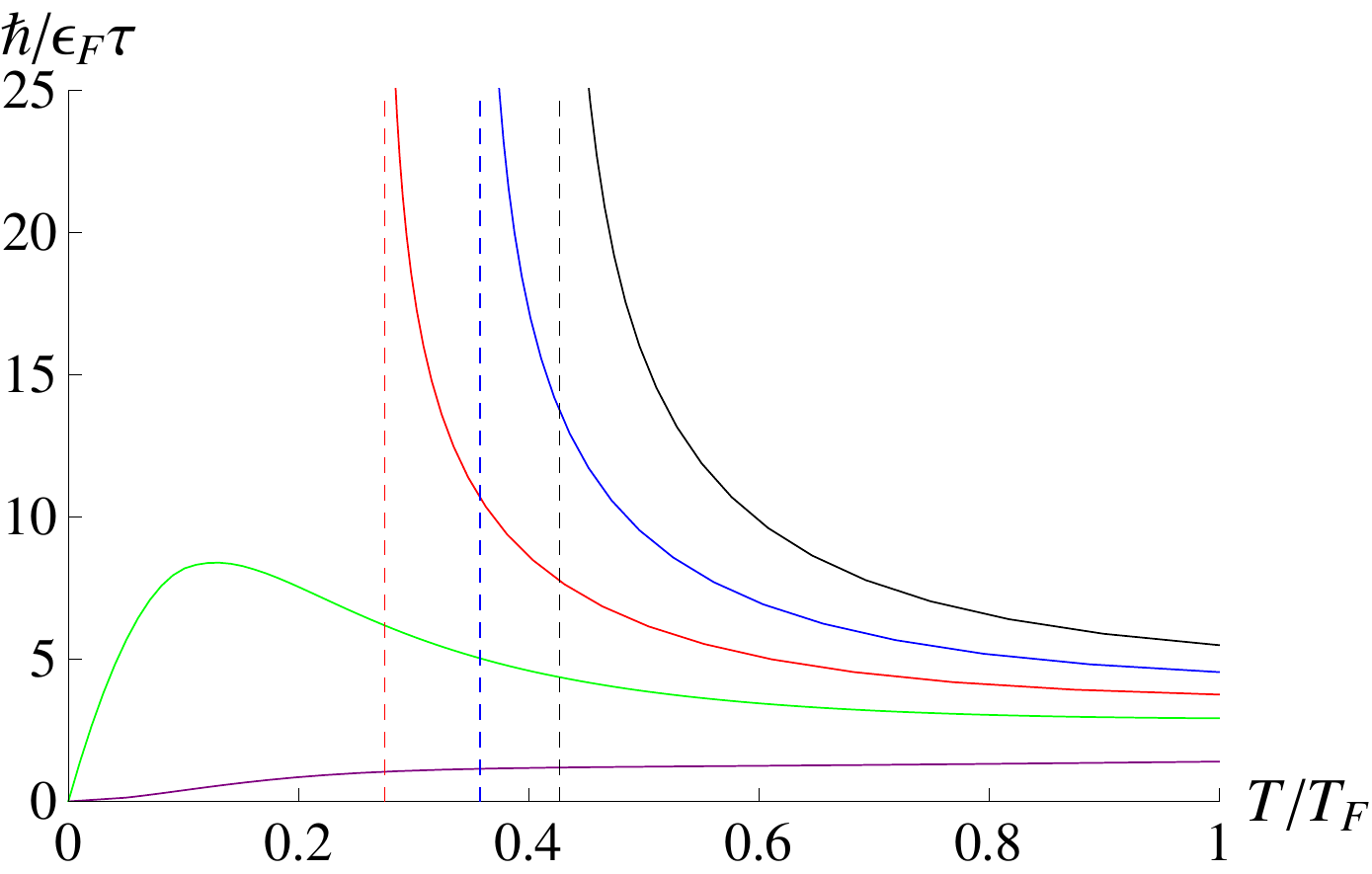}
\caption{\label{fig:mag} (Color online) The dimensionless spin-drag relaxation rate versus temperature in an ultracold Fermi gas with repulsive interactions close to a ferromagnetic transition. From top to bottom the lines correspond to $k_{\rm F} a = 1.9$, $k_{\rm F} a = 1.8$, $k_{\rm F} a = 1.7$, $k_{\rm F} a = \pi/2$, and to $k_{\rm F}a = 1.2$. The critical temperatures corresponding to the $T_{\rm c}$ values for the cases $k_{\rm F} a = 1.7,1.8,1.9$ are indicated by the asymptotes. Note that $T_{\rm c}$ is nonzero only when $k_{\rm F}a>\pi/2$.}
\end{center}
\end{figure}

The result of the complete calculation is shown in Fig.~\ref{fig:mag}, where we plot the dimensionless spin-drag relaxation rate versus temperature. The spin-drag relaxation rate is obtained from the drag resistivity using the Drude formula $\rho_{\rm D} = m/2 n \tau$, where $2n$ is the total density, and $n=k_{\rm F}^3/6\pi^2$ is the density of one spin state. In Fig.~\ref{fig:mag} we see that the spin-drag relaxation rate is enhanced close to the critical temperature and diverges when approaching $T_{\rm c}$ from above. Closer inspection shows that this divergence is logarithmic, as concluded in the previous section.

\subsection{Exciton condensation in an electron-hole bilayer}

As a second application of our formalism we consider an electron-hole bilayer, which can e.g. be obtained in a GaAs-AlGaAs double quantum well. \cite{ehexp} This system was considered previously close to exciton condensation using a diagrammatic approach \cite{hu}. The transition is caused by the formation of pairs (excitons) between electrons in one layer and holes in the other. We consider a system in which the top layer is electron doped with dispersion $\xi_\uparrow({\bm k}) = \hbar^2 k^2/2 m_e - \mu_e$ and the bottom layer is hole doped in which the electrons have the dispersion $\xi_\downarrow({\bm k}) = -\hbar^2 k^2/2 m_h + \mu_h$. In terms of the bare electron mass $m_{\text{bare}}$, the band masses of electrons and holes are $m_e = 0.067 m_{\text{bare}}$ and $m_h = 0.51 m_{\text{bare}}$, respectively.

Close to the transition, the interlayer interaction is determined by the momentum and the energy of an exciton. In the notation of Sec.~\ref{sec:qf} these are ${\bm Q} = {\bm k}_1 - {\bm k}_3$ and $\hb \om = \xi_\uparrow({\bm k}_1) - \xi_\downarrow({\bm k}_3)$, respectively. If we approximate the bare interlayer interaction by a contact interaction, the effective interlayer interaction can be determined to be
\be \label{eq:Weh}
W({\bm k}_1,{\bm k}_2,{\bm k}_3,{\bm k}_4) = W_\text{eff}({\bm Q},\hb \om)  = \frac{V_0}{1+V_0 \De({\bm Q},\hb \om)},
\ee
where the contact interaction strength $V_0 = 4 \pi \hb^2 / m \log(2 \hb^2/ m \mu a^2)$, where $a$ is the two-dimensional scattering length, $\mu$ is the mean chemical potential $\mu = (\mu_e + \mu_h)/2$ and $m$ is the harmonic mean of the electron and hole masses $1/m = 1/2(1/m_e+1/m_h)$. The susceptibility $\Delta({\bm Q},\hbar\omega)$ is
\begin{multline}
\Delta({\bm Q},\hbar\omega) = \frac{1}{V}  \sum_{{\bm k}} \left[\frac{n_{\rm F}(\xi_\uparrow({\bm k}+{\bm Q}))- n_{\rm F}(\xi_\downarrow({\bm k}))}{\xi_\uparrow({\bm k}+{\bm Q}) -  \xi_\downarrow({\bm k})- \hbar \omega - i 0}
\right. \\  \left. -\frac{1}{\ep_\uparrow(\kv) + \ep_\downarrow(\kv) +\mu_\uparrow+\mu_\downarrow}\right],
\end{multline}
where $\ep_\uparrow(\kv) = \hbar^2 k^2/2 m_e$ and $\ep_\downarrow(\kv) = \hbar^2 k^2/2 m_h$.

\begin{figure}
\begin{center}
\includegraphics[width = 0.5 \textwidth]{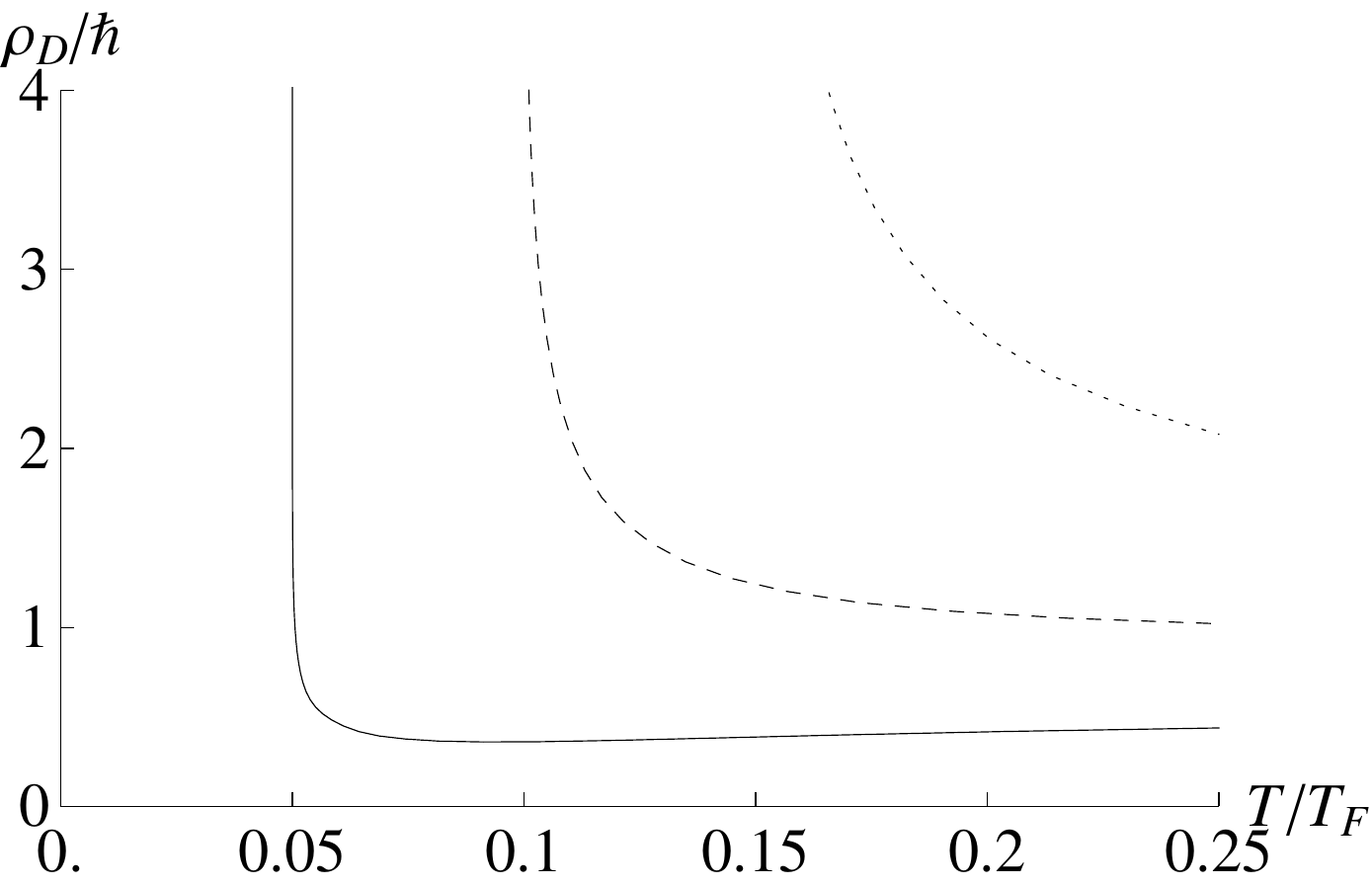}
\caption{\label{fig:EH} The dimensionless drag resistivity $\rho_{\rm D}/\hb$ versus the scaled temperature $T/T_{\rm F}$ for the density balanced case. The solid line corresponds to $k_{\rm F} a = 0.34$ with $T_{\rm c}/T_{\rm F} = 0.05$, the dashed line corresponds to $k_{\rm F} a = 0.68$ with $T_{\rm c}/T_{\rm F} = 0.10$,  and the dotted line corresponds to $k_{\rm F} a = 1.0$ with $T_{\rm c}/T_{\rm F} = 0.15$.}
\end{center}
\end{figure}

To determine the drag resistivity we need to evaluate the collision integral in Eq.~(\ref{eq:gad}) for the scattering amplitude in Eq.~(\ref{eq:Weh}) by decoupling it in the Fock channel, as described in Sec.~\ref{sec:qf}. The result of this calculation is shown in Fig.~\ref{fig:EH}, where we show the dimensionless drag resistivity $\rho_{\rm D}/\hb$ versus the scaled temperature $T/T_{\rm F}$ for the density-balanced case. The solid line corresponds to $k_{\rm F} a = 0.34$ with $T_{\rm c}/T_{\rm F} = 0.05$, the dashed line corresponds to $k_{\rm F} a = 0.68$ with $T_{\rm c}/T_{\rm F} = 0.10$, and the dotted line corresponds to $k_{\rm F} a = 1.0$ with $T_{\rm c}/T_{\rm F} = 0.15$.
As the transition temperature is approached from above, we find that the drag resistivity diverges as $\log(T-T_{\rm c})$, which is in agreement with the theoretical prediction of Ref.~[\onlinecite{hu}] and our qualitative arguments in Sec.~\ref{subsec:qualitative}.

\section{Conclusions}
\label{sec:con}

In summary, we have presented a general formalism to determine the drag resistivity in systems close to a phase transition. We have shown that by decoupling the collision integral in the appropriate channel, we can take into account the effect of Gaussian fluctuations close the phase transition. Our theory is valid for both two and three dimensions, linear and quadratic dispersions, arbitrary mass and population imbalances, and bands of both ``positive" and ``negative" character. The approach presented in this article is valid outside the critical region, by which we mean the region in which the Gaussian approximation breaks down and non-trivial critical exponents become relevant.  In this region critical fluctuations will alter the temperature dependence qualitatively. The width, in temperature, of this region depends on the system at hand. For the case of a ferromagnetic Fermi gas we have checked that the upturn predicted by our Boltzmann theory takes place well outside the critical region. \cite{duinemag} As an application, we determined the spin-drag relaxation rate in an ultracold Fermi gas close to the ferromagnetic transition and the drag resistivity in an electron-hole bilayer close to exciton condensation.

This work was supported by the Stichting voor Fundamenteel Onderzoek der Materie (FOM), the Netherlands Organization for Scientic Research (NWO), and by the European Research Council (ERC). G.V. was supported by NSF grant DMR-1104788.

\section*{APPENDIX}
\label{sec:app}
We show how to take the drift momenta ${\bm k}^\text{drift}$ out of the summation in Eq.~(\ref{eq:gav2})
\begin{multline}
{\bm \Gamma}_\uparrow  = - \frac{2 \pi \beta}{2 \hbar^2 V^3} \sum_{{\bm k}_i}  \delta_{{\bm k}_i} \delta(\xi_i)
|W|^2 n_1 n_2 (1-n_3) (1- n_4)   \\ (\partial_{\bm k} \xi_{4} -\partial_{\bm k} \xi_{1})
\left[
s_\uparrow {\bm k}^\text{drift}_\uparrow \cdot (\partial_{\bm k} \xi_{4}  - \partial_{\bm k} \xi_{1}) +
s_\downarrow {\bm k}^\text{drift}_\downarrow \cdot (\partial_{\bm k} \xi_{3} - \partial_{\bm k} \xi_{2})
\right].
\end{multline}
For a quadratic dispersion this is easy, since $\partial_{\bm k} \xi_{i} \propto {\bm k}_i$. Using  momentum conservation ${\bm k}_1 + {\bm k}_2 = {\bm k}_3 + {\bm k}_4$ and that
\be
\int d {\bm q} F(q) {\bm q} ({\bm q} \cdot {\bm k}^\text{drift}) = \frac{{\bm k}^\text{drift}}{d} \int d {\bm q} F(q) {\bm q}^2,
\ee
where $F(q)$ is a function that only depends on the length of $q$, immediately leads to the results Eqs.~(\ref{eq:gasu},\ref{eq:gasd},\ref{eq:gad}). When we have a two-dimensional system with linear dispersion, then $\partial_{\bm k} \xi \propto \hat{{\bm k}}$. The collision integral ${\bm \Gamma}_\uparrow$ in Eq.~(\ref{eq:gav2}) is invariant under a global rotation of all integration vectors, in particular, operating with
\be
\frac{1}{2 \pi} \int d \phi
\ee
where the angle $\phi$ is a global rotation angle around an $z$-axis of all vectors ${\bm k}_i$, leaves the integral invariant. The $\delta$'s, the interaction and the Fermi functions do not depend on $\phi$. We write the integral for a general $\phi$-dependent term
\begin{multline}
\frac{1}{2 \pi} \int d \phi {\bm k}_a ({\bm k}^\text{drift} \cdot {\bm k}_b) = \frac{k_a k_b}{2 \pi} \int d \phi
\begin{pmatrix}
\cos(\phi + \phi_a) \\ \sin(\phi + \phi_a)
\end{pmatrix} \\
\left[
\begin{pmatrix}
k^\text{drift}_x \\ k^\text{drift}_y
\end{pmatrix}
\cdot
\begin{pmatrix}
\cos(\phi + \phi_b) \\ \sin(\phi + \phi_b)
\end{pmatrix}
\right],
\end{multline}
where ${\bm k}_a$ and ${\bm k}_b$ are momenta being summed in Eq.~(\ref{eq:gav2}), and
we choose the $x$-axis along the external drift momentum ${\bm k}^\text{drift}$, where ${\bm k}_a$ and ${\bm k}_b$ can either be the same or different. Now, reflecting the whole integrand in the $x$-axis and then doing the integral gives the same answer. This sends all angles $\phi_i$ to $-\phi_i$. Again, the $\delta$'s, interaction, and Fermi functions do not depend on $\phi$ and the sign of the angles, so we get the term
\begin{multline}
\frac{1}{2 \pi} \int d \phi {\bm k}_a ({\bm k}^\text{drift} \cdot {\bm k}_b) = \frac{k_a k_b}{2 \pi} \int d \phi
\begin{pmatrix}
\cos(\phi - \phi_a) \\ \sin(\phi - \phi_a)
\end{pmatrix} \\
\left[
\begin{pmatrix}
k^\text{drift}_x \\ k^\text{drift}_y
\end{pmatrix}
\cdot
\begin{pmatrix}
\cos(\phi - \phi_b) \\ \sin(\phi - \phi_b)
\end{pmatrix}
\right]\,.
\end{multline}
Now, we can take the integrals over $\phi$ and perform the average of the two terms, leading to the replacement rule
\be
{\bm k}_a ({\bm k}^\text{drift} \cdot {\bm k}_b) \quad \rightarrow \quad \frac{1}{2} {\bm k}^\text{drift} ({\bm k}_a \cdot {\bm k}_b),
\ee
which can be used to obtain Eqs.~(\ref{eq:gasu},\ref{eq:gasd},\ref{eq:gad}). Note that for the special case ${\bm k}_a = {\bm k}_b$, we recover the result for used for the quadratic dispersion.

\end{document}